\documentclass[aps,prd,floatfix,nofootinbib,superscriptaddress,twocolumn,showkeys]{revtex4-2}
\usepackage{amsmath}
\usepackage{amssymb}
\allowdisplaybreaks

\usepackage[thicklines]{cancel}
\usepackage{makeidx}
\usepackage{amsfonts}
\usepackage[usenames,dvipsnames]{pstricks}
\usepackage{subfigure}
\usepackage{epsfig}
\usepackage{pst-grad}
\usepackage{mathtools}
\usepackage[colorlinks,hyperindex]{hyperref}
\usepackage{array}
\usepackage{tikz}
\usepackage{float}
\usetikzlibrary{decorations.markings,math,shapes.misc,patterns,arrows.meta,calc,patterns,angles,quotes,intersections}
\usepackage{allpurpose}
\newcommand {\nn}{\nonumber}
\usepackage{mathrsfs}
\usepackage{wasysym}
\allowdisplaybreaks[4]
\usepackage{tabularx}
\usepackage{longtable}
\hypersetup
{colorlinks,
citecolor=black,
linkcolor=black,
urlcolor=black,
}

\begin{document}

\title{Periapsis shift in spherically symmetric spacetimes and effect of electric interaction}

\author{Qianchuan Wang}
\address{School of Physics and Technology, Wuhan University, Wuhan, 430072, China}

\author{Junji Jia}
\email[Corresponding author:~]{junjijia@whu.edu.cn}
\address{Department of Astronomy \& MOE Key Laboratory of Artificial Micro- and Nano-structures, School of Physics and Technology, Wuhan University, Wuhan, 430072, China}

\date{\today} 

\begin{abstract}
The periapsis shift of charged test particles in arbitrary static and spherically symmetric charged spacetimes are studied. Two perturbative methods, the near-circular approximation and post-Newtonian methods, are developed, and shown to be very accurate when the results are found to high orders. The former method is more precise when the eccentricity $e$ of the orbit is small while the latter works better when the orbit semilatus rectum $p$ is large. Results from these two methods are shown to agree with each other when both $e$ is small and $p$ is large. These results are then applied to  the Reissner-Nordstr\"om spacetime, the Einstein-Maxwell-dilation gravity and a charged wormhole spacetime. The effects of various parameters on the periapsis shift, especially that of the electrostatic interaction,  are carefully studied. The periapsis shift data of the solar-Mercury is then used to constrain the charges of the Sun and Mercury, and the data of the Sgr A$^*$-S2 periapsis shift is used to find, for the first time using this method, constraints about the charges of Sgr A$^*$ and S2.
\end{abstract}


\maketitle

\section{Introduction}

The periapsis shift (PS) of celestial bodies around the central more massive gravity source, is an important experimental and theoretical tool for the investigation of gravity and the properties of celestial bodies and their orbits. 
Besides the famous example of Einstein's explanation of Mercury's extra PS using General Relativity (GR) \cite{einstein1916}, nowadays PS of different kinds of objects such as satellites \cite{lucchesi2011,Everitt:2011hp}, planets in the solar system \cite{Iorio:2018adf}, and stars around Sgr A$^*$ \cite{GRAVITY:2020gka} can be observed experimentally. 
The PS of these objects are related not only to the properties of the test particles (such as their charge, spin or weak interaction with the environment \cite{Iorio:2018adf}) but also those of the central object (charge, spin, oblateness etc. \cite{Iorio:2018adf,GRAVITY:2020gka, He:2023joa}), and the nature of the gravity itself \cite{Harko:2011kv}. Moreover, the PS is also dependent on the kinetic parameters of the orbits, such as their effective eccentricity $e$ or semilactus rectum $p$, or 
equivalently the specific energy $E$ and angular momentum $L$. 

Among the intrinsic parameters of the spacetime, the central object mass $M$'s effect on the PS is the most well-studied and the result is the well-known Schwarzschild one \cite{einstein1916}. The next frequently considered parameter, whose effect on the PS is also observationally important, is the spin angular momentum of the central object, especially that of those more compact sources. 
The effect of the electric charge $Q$ of the source on the PS, which is the first motivation of the current work, is less commonly investigated. This parameter is expected to affect the PS of test particles 
gravitationally through the gravielectric effect. 
Previously, it was known that this effect will always decrease the deflection angle of test particle trajectories in the weak field limit, regardless of the sign of $Q$ \cite{Pang:2018jpm,Zhou:2022dze}. 
This implies that a nonzero $Q$ would also decrease the PS of test particles. 
One of the main purposes of this work is to show that this is indeed the case, at least to the leading order of the post-Newtonian (PN) approximation, for all kinds of electrically charged spacetimes. 

A more interesting case is when the test particle is also charged with $q$, and consequently, both the gravielectric and the true electrical effects will affect the PS. Moreover, due to the fact that the electric interaction strength is much stronger than the gravitational one, the charge effect can still be non-negligible even when both the spacetime and particle charges are small. For the electric effect, the sign of $Q$ matters because its product with $q$, i.e. sign($qQ$) determines the nature of electric interaction, and therefore $Q$ could potentially increase the PS too. This means that there is an interplay between the attractive gravitational interaction and the (potentially repulsive) Coulomb interaction. How these interactions and parameters $(Q,\,q)$ would affect the PS is unclear and this forms the second motivation of this work. 

Previously, when finding the PS under pure gravity, most of the works used essentially some kind of PN approximation, i.e., they assumed that the trajectory is far from the center $(r\gg M)$ and close to an ellipse \cite{Avalos-Vargas:2011jja, Mak:2018hcn, Shchigolev:2015sgg,Heydari-Fard:2019pxd, Gong:2009zzb, Gong:2008zzb, Heydari-Fard:2019kno, Hu:2013eya, Zhang:2022zox, Yang:2019aez}.  However, when the trajectory is close to the center, especially when the center is a black hole such as Sgr A$^*$, high order results which were hard to obtain using the PN approximation, might be needed to meet the required accuracy. In this work, we will develop a near-circular (NC) approximation that works for orbits that are not very large. 

For the PN method, most of the previous calculations only worked in specific spacetimes to the leading or at most next-to-leading order of $M/p$ \cite{Avalos-Vargas:2012jja, Bakry:2021smq, Jiang:2014kxz, Yang:2022wve, Yang:2022sco, Heydari-Fard:2019pxd}. In this work, we will also systematically develop the PN method in general static and spherically symmetric spacetimes with nonzero charges, in such a way that not only the gravitational but electric interactions can be treated simultaneously, and the PS to very high order will be obtained and linked directly to the asymptotic expansions of the metric functions of the spacetime. 
 
The paper is organized as follows. In Sec. \ref{sec:method1}, we lay out the preliminaries, the general form of the spacetime and electric field, the definition of the PS, the NC approximation method and the corresponding result $\alpha_{\mathrm{NC}}$. Sec. \ref{sec:NTmethod} is devoted to the PN treatment of the PS and the result $\alpha_{\mathrm{PN}}$ to arbitrary order. The equivalence of the two approximations are also shown here. In Sec. \ref{sec:applications}, these two methods are applied to the Reissner-Nordstr\"om (RN) spacetime, the Einstein-Maxwell dilaton (EMD) gravity and the charged wormhole spacetimes. The PS in each of them is found and the effect of the electric interaction is carefully analyzed. Sec. \ref{sec:appdata} is devoted to the constraining of the charge of the solar-Mercury system and the Sgr A$^*$-S2 system using the data of the PS in each system. Sec. \ref{sec:disc} concludes the work with a short discussion. Throughout the work, we use the natural system of units $G=c=4\pi\epsilon_0=1$. 

\section{Preliminaries and the NC approximation\label{sec:method1}}

We study the motion of a charged test particle in the most general static and spherically symmetric spacetime whose line element is described by
\begin{equation}
    \mathrm{d}s^2=-A(r)\mathrm{d}t^2+B(r)\mathrm{d}r^2+C(r)(\mathrm{d}\theta^2+\sin^2\theta\mathrm{d}\phi^2), 
    \label{eq:gmetric}
\end{equation}
where $(t,r,\theta,\phi)$ are the coordinates and the metric functions depend on $r$ only. For the electromagnetic field in this spacetime, we will assume that there is only a pure electric field described by the potential $\left[\mathcal{A}_0\left(r\right),0,0,0\right]$. In some spacetime solutions with electric fields, however, no such potential but the electric field itself $E(r)$ was given. Fortunately, for the two methods presented in this paper, what is needed is only the series expansion of $\mathcal{A}_0(r)$ near the orbit radius and therefore the electric field $E(r)$ is also enough.
Due to the symmetry of the spacetime and field, without losing any generality, we can always investigate the equatorial motion of charged test particles in such spacetimes. 

The dynamics of a point particle possessing a non-zero mass $m$ and specific charge $\hat{q}\equiv q/m$ can be described by the Lorentz equation \cite{Tursunov:2018erf}
\be\label{eq:Lorentz}
\frac{\dd ^2x^\rho}{\dd \tau^2}+\Gamma^{\rho}{}_{\mu\nu}\frac{\dd x^\mu}{\dd \tau}\frac{\dd x^\nu}{\dd \tau}=\hat{q}\mathcal{F} ^{\rho}{}_{\mu}\frac{\dd x^\mu}{\dd \tau},
\ee
where $F_{\mu\nu}=\partial_\mu A_\nu-\partial_\nu A_\mu$ and $\tau$ is the proper time of the test particle. 
After integrating Eq. (\ref{eq:Lorentz}) on the equatorial plane $(\theta=\frac{\pi}{2})$, we obtain three first-order differential equations \cite{Zhou:2022dze} 
\begin{subequations}\label{eqs:motion}
\begin{align}
    \dot{t}=&\frac{\hat{E}+\hat{q}\mathcal{A}_0}{A},\\
    \dot{\phi}=&\dfrac{\hat{L}}{C},\\ 
    \dot{r}^2=&\left[\frac{(\hat{E}+\hat{q}\mathcal{A}_0)^2}{AB}-\frac{1}{B}\right]-\frac{\hat{L}^2}{BC},\label{eq:motion_r}
\end{align}
\end{subequations}
where the integration constants $\hat{E}$ and $\hat{L}$ correspond to the energy and angular momentum per unit mass of the particle respectively, and $\dot{{}}$ denotes the derivative with respect to the proper time $\tau$. 

To analyze the radial motion of the particle in more detail, we first rewrite Eq. (\ref{eq:motion_r}) in an alternative form 
\be
 B(r)C(r)\dot{r}^2=\left[\frac{(\hat{E}+\hat{q}\mathcal{A}_0(r))^2}{A(r)}-1\right]C(r)-\hat{L}^2. 
\label{eq:effapp}
\ee
The zeros and singularities of $B(r)$ and $C(r)$ generally correspond to event horizons or singularities of the spacetime, and in this work we will concentrate on the region $B(r)C(r)>0$. Then we can think of the first term on the right-hand side of Eq. \eqref{eq:effapp} as some kind of effective potential, against which $\hat{L}^2$ can be compared with
\begin{equation}\label{eq:VL}
V(r)=\left[\frac{(\hat{E}+\hat{q}\mathcal{A}_0(r))^2}{A(r)}-1\right]C(r).
\end{equation}
Then the motion of the particle is only allowed when $V(r)\geq \hat{L}^2$.
We assume that this inequality has a nonempty interval solution $[r_1,~r_2]$ where 
\be\label{eq:r1r2}
V(r_1)=V(r_2)=\hat{L}^2.
\ee
Clearly, $r_1$ and $r_2$ are the radii of the periapsis and apoapsis.
For simplicity, we will also assume that there is only one local maximum of $V(r)$ at $r=r_c$ in this interval. For later usage, we will denote the value $V(r_c)$ as $\hat{L}_c^2$, i.e., 
\be \label{eq:rc}
\frac{\dd V(r)}{\dd r}\bigg|_{r=r_c}=0,~V(r_c)\equiv \hat{L}_c^2. 
\ee
The existence of such $r_c$ is always possible as long as $\hat{L}^2$ is close enough to $\hat{L}_c^2$. The physical interpretation of $r_c$ is straightforward: it represents the radius of the circular orbit of the particle with energy $\hat{E}$ and angular momentum $\hat{L}_c$. We point out that except for very simple spacetime and electric potential (e.g., the RN spacetime), it is usually difficult to solve the algebraic equation \eqref{eq:rc} to obtain the closed form solution of $r_c$. Therefore in the following sections where $r_c$ is used, we will have to assume that $r_c$ is already solved, numerically if necessary.  

With the above consideration, the PS can be defined as 
\be\label{eq:alpha_raw_raw}
\begin{aligned}
\alpha=&2\int_{r_1}^{r_2}\left|\frac{\dd \phi}{\dd r}\right|\dd r -2\pi,
\end{aligned}
\ee
which after using Eqs. (\ref{eqs:motion}) becomes
\be\label{eq:precession_defination_raw}
\begin{aligned}
\alpha=2\left(\int_{r_1}^{r_c}+\int_{r_c}^{r_2}\right)\sqrt{\frac{B(r)}{C(r)}} \frac{\hat{L} \mathrm{~d} r}{\sqrt{V(r)-\hat{L}^2}}-2\pi.
\end{aligned}
\ee
In order to reveal the effect of electric interaction on the PS, our task becomes to systematically solve the PS in Eq. (\ref{eq:precession_defination_raw}) using appropriate approximations. In this section, we will present a perturbative approach based on the NC approximation, and then in the next section, a PN method will be used.

For the NC approximation, the first step for the computation of the integrals in Eq. \eqref{eq:precession_defination_raw} is to utilize the deviation of the particle's orbit from a perfect circle as a measure of perturbation for series expansion. This deviation can be quantified by the parameter $a$ defined as
\be\label{eq:a}
a\equiv\frac{\hat{L}_c}{\hat{L}}-1.
\ee
When $r_1$ and $r_2$ are close, i.e., the orbit is close to a circle, then the above defined $a$ will be a small quantity. 

In order to use $a$, we will extend the method developed in Ref. \cite{Zhou:2022dze,Jia:2020qzt,He:2023joa} to the case with electric interaction here. We first propose a change of variables in Eq. (\ref{eq:precession_defination_raw}) 
from $r$ to $\xi$, which are linked by
\be
\xi\equiv \hat{L}_c P\left(\frac{1}{r}\right), \label{eq:xi}
\ee
where the function $P(1/r)$ is defined as
\be\label{eq:P}
P\left(\frac{1}{r}\right)\equiv \frac{1}{\sqrt{V(r)}}-\frac{1}{\hat{L}_c}.
\ee
From Eq. (\ref{eq:rc}), we observe that the function $P(x)$ exhibits opposite monotonic behavior on the two sides of $x=1/r_c$. This implies that its inverse function possesses two different branches in the domain of $x\geqslant 0$. We use $\omega_+$ and $\omega_-$ to denote the branch of the inverse function of $P(x)$ for $r>r_c$ and $r< r_c$ respectively. In other words, using \eqref{eq:xi} we have
\be
\label{eqs:integrand_trans}
r= \left\{\begin{aligned}
        &1/\omega_{-}\left( \xi/\hat{L}_c\right)\ \text{for}~ r<r_c,\\
        &1/\omega_{+}\left( \xi/\hat{L}_c\right)\ \text{for}~ r\geqslant r_c.
    \end{aligned}\right.
\ee
Substituting this change of variables, the terms in the integral (\ref{eq:precession_defination_raw}) become
\begin{subequations}
\begin{align}
&\frac{V(r)}{\hat{L}^2}-1 \to \frac{(2+a+\xi)(a-\xi)}{(\xi+1)^2},\\
    \label{eqs:interval_trans}
    &r_{1,2}  \to \hat{L}_cP\left(\frac{1}{r_{1,2}}\right)=a,\\
    &r_c   \to \hat{L}_cP\left(\frac{1}{r_{c}}\right)=0,
\end{align}
\end{subequations}
and consequently the PS in Eq. (\ref{eq:precession_defination_raw}) can be rewritten as 
\be\label{eq:alpha_re}
\alpha=\int_0^a \frac{y(a,\,\xi)}{\sqrt{a-\xi}} \mathrm{~d} \xi -2\pi 
\ee
where the factor $y(a,\,\xi)$ of the integrand is
\begin{align}\label{eq:yxi_raw}
y(a,\,\xi)=    2\sum_{s=+,-}s\sqrt{\frac{B(1 / \omega_s)}{C(1 / \omega_s)}} \frac{\xi+1}{\hat{L}_c \sqrt{2+a+\xi}} \frac{-\omega_s^{\prime}}{\omega_s^2}.
\end{align}
An evident advantage of this change of variables is that it transforms the two different integral limits $r_1$ and $r_2$ into the same value $a$ and therefore simplifies the computation. 

Since we are concentrating on the $a\ll 1$ case, the next step is naturally expanding the function $y(\xi)$ for small $\xi$. Carrying out this expansion, one can show that in general it always takes the form
\be\label{eq:ykraw}
\begin{aligned}
y(a,\,\xi)=\sum_{n=0}^{\infty}y_n(a) \xi^{n-\frac12}.
\end{aligned}
\ee
Here the coefficients $y_n(a)$ can be determined once the metric functions and potential $\mathcal{A}_0$ are known around $r_c$. 
Substituting this back into Eq. \eqref{eq:alpha_re}, and using the integral formula
\be\label{eq:intF}
\int_0^a \frac{\xi^{n-\frac12}}{\sqrt{a-\xi}}\mathrm{~d}\xi=\frac{(2n-1)!!}{(2n)!!}\pi a^n, 
\ee
the PS is computed as
\begin{align}
\label{eq:alpha_order_n}
\alpha=&\sum_{n=0}^{\infty}\frac{(2n-1)!!}{(2n)!!}\pi y_n(a) a^n-2\pi.
\end{align}
Since we have assumed that the deviation parameter $a$ is small, then in principle we can further expand $y_n(a)$ for small $a$ and reorganize the PS into a true power series
\begin{align}
\alpha_{\mathrm{NC}}=&\sum_{n=0}^{\infty}t_n a^n =\sum_{n=0}^{\infty}t_n \lb\frac{\hat{L}_c}{\hat{L}}-1\rb^n .\label{eq:alphaint}
\end{align}
These coefficients $t_n$ are now independent of $a$ and, similar to $y_n$, can be completely determined once the metric functions and electric potential are fixed. We now show this determination process in the following. 

We assume that the metric functions and the electric potential $\mathcal{A}_0$ have the following expansions around $r_c$
\begin{subequations}\label{eq:abcform1}
\begin{align}
&A(r)=\sum_{n=0}^{\infty}a_{cn}(r-r_c)^n,\\
&B(r)=\sum_{n=0}^{\infty}b_{cn}(r-r_c)^n,\\
&C(r)=\sum_{n=0}^{\infty}c_{cn}(r-r_c)^n,\\
&\mathcal{A}_0(r)=\sum_{n=0}^{\infty}h_{cn}(r-r_c)^n.
\end{align}
\end{subequations}
The definition of $r_c$ in Eq. (\ref{eq:rc}) immediately implies a constraint between these coefficients
\begin{align}\label{eq:EonRc}
\hat{E}+\hat{q}h_{c0}=&\frac{a_{c0}c_{c0}}{a_{c0}c_{c1}-a_{c1}c_{c0}}\nn\\
&\times \left(\sqrt{\frac{c_{c1}}{c_{c0}^2}\left(a_{c0}c_{c1}-a_{c1}c_{c0}\right)+\hat{q}^2h_{c1}^2}-\hat{q}h_{c1}\right).
\end{align}
We will use this relation occasionally to simplify the $t_n$ and $T_n$ below. The critical $\hat{L}_c$ in Eq. \eqref{eq:alphaint} can also be expressed in terms of these coefficients using again Eq. \eqref{eq:rc}
\be
\hat{L}_c=\sqrt{c_{c0} \left[\frac{\left(\hat{E}+\hat{q} h_{c0}\right){}^2}{a_{c0}}-1\right]}. 
\ee

Substituting the expansions in Eqs. \eqref{eq:abcform1} into Eqs. \eqref{eq:P}, and taking its inverse function $\omega_\pm$ and then substituting into \eqref{eq:yxi_raw} and carrying out the expansion for small $\xi$ and eventually expanding $y_n(a)$ again for small $a$, one finds the coefficients $t_n$ in Eq. \eqref{eq:alphaint}. Here for simplicity, we only list the first two of them
\begin{widetext}
\begin{subequations}\label{eqs:tn}
\begin{align}
    t_0=& \frac{ 2\pi b_{c0}^{1/2} \hat{L}_c }{c_{c0}^{1/2}\left(-T_2\right)^{1/2}},\\
    t_1=& -\frac{\pi }{4 b_{c0}^{3/2} c_{c0}^{5/2} (-T_2)^{7/2}}\left\{\left[\left(-3 b_{c0}^2 c_{c1}^2+4 b_{c0}^2 c_{c0} c_{c2}-4 b_{c2} b_{c0} c_{c0}^2+2 b_{c1} b_{c0} c_{c0} c_{c1}+b_1^2 c_{c0}^2\right)  T_2^2\right.\right.\nn\\
    &\left.\left.+\left(6 b_{c0} b_{c1} c_{c0}^2-6 b_{c0}^2 c_{c0} c_{c1}\right) T_2 T_3 -15 b_{c0}^2 c_{c0}^2  T_3^2+12 b_{c0}^2 c_{c0}^2  T_2 T_4\right]\hat{L}_c^3 -8 b_{c0}^2 c_{c0}^2 T_2^3  \hat{L}_c \right\},
\end{align}
\end{subequations}
where the $T_n$ are the $n$-th derivative of the potential at $r_c$, i.e., $T_n\equiv V^{(n)}(r_c)/n!$ and the first two of them are
\begin{subequations}
\begin{align}
T_2=&\frac{\left(\hat{E}+\hat{q}h_{c0}\right)^2}{a_{c0}^3}\left(a_{c0}^2 c_{c2}-a_{c2} a_{c0} c_{c0}-a_{c1} a_{c0} c_{c1}+a_{c1}^2 c_{c0}\right)\nn\\
&+\frac{2 \hat{q} \left(\hat{E}+\hat{q}h_{c0}\right)}{a_{c0}^2}\left(-a_{c1} c_{c0} h_{c1}+a_{c0} c_{c1} h_{c1}+a_{c0} c_{c0} h_{c2}\right)+\frac{ \hat{q}^2 c_{c0} h_{c1}^2}{a_{c0}}-c_{c2} ,\\
T_3=&\frac{\left(\hat{E}+\hat{q}h_{c0}\right)^2}{a_{c0}^4}\left[a_{c0} a_{c1}^2 c_{c1}+a_{c0} a_{c1} \left(2 a_{c2} c_{c0}-a_{c0} c_{c2}\right)-a_{c1}^3 c_{c0}+a_{c0}^2 \left(-a_{c3} c_{c0}-a_{c2} c_{c1}+a_{c0} c_3\right)\right]\nn\\
&+\frac{2 \hat{q} \left(\hat{E}+\hat{q}h_{c0}\right)}{a_{c0}^3}\left[a_{c0} a_{c0} \left(c_{c2} h_{c1}+c_{c1} h_{c2}+c_{c0} h_{c3}\right)-a_{c0} a_{c2} c_{c0} h_{c1}-a_{c0} a_{c1} \left(c_{c1} h_{c1}+c_{c0} h_{c2}\right)\right.\nn\\
&\left.+a_{c1}^2 c_{c0} h_{c1}\right]+\frac{\hat{q}^2 h_{c1} }{a_{c0}^2}\left(-a_{c1} c_{c0} h_{c1}+a_{c0} c_{c1} h_{c1}+2 a_{c0} c_{c0} h_{c2}\right)-c_{c3} .
\end{align}
\end{subequations}
\end{widetext}
The higher order $t_n~(n=2,\,3,\cdots)$ and  $T_n~(n=4,\,5,\cdots)$ can also be obtained easily with the help of an algebraic system. 

In obtaining the PS in Eq. \eqref{eq:alphaint}, although we used the small $a$ or equivalently the small eccentricity $e$ approximation, we do not have to assume that the orbit size itself is large compared to the characteristic length scale (e.g. mass $M$) of the spacetime. The latter is the key assumption for the PN method, which can also be used to compute the corresponding PS. In next section, we show how the PN method can be used to take into account the electric interaction and to find the PS when the orbit size is large. 

\section{PS using PN method\label{sec:NTmethod}}

Many previous works computed the PS using the PN methods. However most of them, if not all, only focused on the lowest order(s) result, and some used this method very loosely and the results are erroneous. Moreover, only a few of them have attempted the PN method to deal with the gravitational and electric interactions simultaneously. In this section, we will systematically develop the PN method that not only yields the PS to high order of the semilatus rectum $p$ but takes into account the electric interaction too. Some of the techniques we used here are similar to those in Ref.  \cite{yongjiu:1990,Gong:2009zzb,He:2023joa}. We will further show that the PN method result agrees perfectly with the result obtained in last section using the NC method, when both the orbit size is large and eccentricity is small. 

For the PN method, we first assume that the orbit can be described by the relation
\begin{equation}\label{eq:FuOrbit}
\dfrac{1}{r}=\dfrac{1}{p}\left[1+e\cos \psi\left(\phi\right)\right]
\end{equation}
where the function $\psi(\phi)$ describes the deviation of the orbit from ellipse. 
Note that the periapsis and apoapsis of the orbit correspond to $\psi=0$ and $\psi=\pi$ respectively, and the radii of the periapsis $r_1$ and apoapsis $r_2$ satisfy
\begin{equation}\label{r12OnP}
    \frac{1}{r_1}=\frac{1+e}{p},\quad \frac{1}{r_2}=\frac{1-e}{p}.
\end{equation}
The radius of the orbit evolves for a full period when $\psi$ changes from 0 to $2\pi$ and therefore the PS using this description becomes
\begin{equation}\label{eq:alphaVraw}
\begin{aligned}
    \alpha_{\mathrm{PN}}= \int_0^{2\pi}\frac{\dd \phi}{\dd \psi}\dd \psi-2\pi .
\end{aligned}
\end{equation}
The $\dd\phi/\dd\psi$ can be expressed from Eq. \eqref{eq:FuOrbit} as
\be\label{eq:dudphi}
\frac{\dd \phi}{\dd \psi}=-\frac{e}{p} \frac{\sin \psi}{\dd u / \dd \phi} ,
\ee
whence we have defined $u=1/r$.
The term $\dd u/\dd\phi$ in Eq. \eqref{eq:dudphi} can be transformed using Eq. (\ref{eqs:motion}) to
\begin{align}\label{eq:FuSeries}
&\left(\frac{\dd u}{\dd \phi}\right)^2 =u^4\frac{\dot{r}^2}{\dot{\phi}^2}\nn\\
=&u^4\left[\frac{(\hat{E}+\hat{q}\mathcal{A}_0(\frac{1}{u}))^2C(\frac{1}{u})^2}{A(\frac{1}{u})B(\frac{1}{u})\hat{L}^2}-\frac{C(\frac{1}{u})^2}{B(\frac{1}{u})\hat{L}^2}\right]-\frac{u^4C(\frac{1}{u})}{B(\frac{1}{u})}\nn\\
\equiv &F(u)
\end{align}
where we have defined the right-hand side as $F(u)$.
Next we will show that $F(u)$ can be expressed as a serial function of $\psi$ and after substituting back into \eqref{eq:alphaVraw}, the integral over $\psi$ can be carried out to find the PS.

The key to accomplishing this is the PN assumption, i.e., we will assume that $p$ is large and therefore $u$ is always a small quantity. This allows us to expand $F(u)$ as a power series of $u$
\be
F(u)=\sum_{n=0}^\infty f_nu^n, \label{eq:cfuexp}
\ee
where the coefficients $f_n$ can be determined from Eq. \eqref{eq:FuSeries} using the asymptotic expansion of the metric functions and $\mathcal{A}_0$. 
Assuming that they are of the form
\begin{align}
A(r) =1+\sum_{n=1} \frac{a_n}{r^n},&\ B(r) =1+\sum_{n=1} \frac{b_n}{r^n},\nn\\
\frac{C(r)}{r^2}=1+\sum_{n=1} \frac{c_n}{r^n},&\ \mathcal{A}_0 =\sum_{n=1} \frac{\mathfrak{q}_n}{r^n}\label{eq:abcform2},
\end{align}
we then are able to find $f_n$ order by order as functions of the coefficients $a_n,~b_n,~c_n$ and $\mathfrak{q}_n$. Here we only illustrate the first few of them as
\begin{subequations}\label{eqs:FuCoefficients}
\begin{align}
f_0=&\frac{ \hat{E}^2-1}{ \hat{L}^2},\\
f_1=&\frac{  \left(b_1 -2  c_1\right) \left( 1-\hat{E}^2\right)- a_1 \hat{E}^2-2 \hat{q} \mathfrak{q}_1 \hat{E}}{\hat{L}^2},\\
f_2=&-1+\frac{1}{\hat{L}^2}\left\{\left(a_1 b_1-2 a_1 c_1+a_1^2-a_2\right)\hat{E}^2\right.\nn\\
&\left.+\left(-2 b_1 c_1+b_1^2-b_2+c_1^2+2 c_2\right) \left(\hat{E}^2-1\right)\right.\nn\\
&\left.+ 2 \hat{q} \hat{E}\left[ \left(-a_1-b_1+2 c_1\right)\mathfrak{q}_1 +\mathfrak{q}_2\right]  + \hat{q}^2 \mathfrak{q}_1^2\right\}.
\end{align}
\end{subequations}

Repetitively substituting Eq. \eqref{eq:FuOrbit} as $u$, all $u$ in the right hand side of Eq. \eqref{eq:cfuexp} can be completely replaced by variables $p,\, e$ and $\psi$. After collecting terms proportional to $\cos\psi$ and $\sin^{2n}\psi$, we are able to show that $F(u)$ always takes the form \cite{He:2023joa}
\begin{align}\label{eq:FuCompare}
F(u)=&G_0(\hat{E},\hat{L},p,e)+G_0^\prime(\hat{E},\hat{L},p,e)\cos \psi\nn\\
&+\sum_{n=1}^\infty G_n(\hat{E},\hat{L},p,e,\cos \psi)\sin^{2n} \psi
\end{align}
where $G_0,~G_0^\prime,~G_n$ are linear combinations of $f_n$ with coefficients being power series functions of $\hat{E},~1/\hat{L},~e$ and $1/p$. 
Their exact forms can be worked out without difficulty, although the algebra is too tedious to show here. 

Since at the apoapsis and periapsis, by definition \eqref{eq:FuSeries} we have $F(u)=0$, using the condition \eqref{r12OnP}, we immediately have $F(\psi=0)=F(\psi=\pi)=0$, i.e., 
\begin{align}\label{G0G1}    G_0(\hat{E},\hat{L},p,e)=0,~G_0^\prime(\hat{E},\hat{L},p,e)=0.
\end{align}
These two power series conditions effectively establish two relations between the kinetic variables $(\hat{E},\, \hat{L})$ and $(p,\, e)$. These conditions can be solved perturbatively when $p$ is large, allowing us to express $(\hat{E},\, \hat{L})$ in terms of $(p,\, e)$
\begin{subequations}\label{eq:NewTonEL}
    \begin{align}
        &\hat{E}=1-\frac{\left(1-e ^2\right) \left(2 \hat{q} \mathfrak{q}_1 -a_1 \right)}{4 }\frac{1}{p}\nn\\
        &+\frac{\left(1-e ^2\right)^2 \left(2 \hat{q} \mathfrak{q}_1 -a_1 \right) \left(2 \hat{q}  \mathfrak{q}_1 -3 a_1 +4 c_1 \right)}{32  }\frac{1}{p^2}+\mathcal{O}\left(p\right)^3,\\
        &\hat{L}^2=\frac{ m \left(2 \hat{q} \mathfrak{q}_1  -a_1 \right)}{2 }p\nn\\
        &+\frac{1}{4 }\left\{\left[-3 a_1 c_1+3 a_1^2-4 a_2+e^2 \left(a_1^2-a_1 c_1\right)\right]\right.\nn\\
        &\left.+\hat{q} \left[-5 a_1 \mathfrak{q}_1+6 c_1 \mathfrak{q}_1+8 \mathfrak{q}_2+e^2 \left(2 c_1 \mathfrak{q}_1-3 a_1 \mathfrak{q}_1\right)\right]\right.\nn\\
        &\left.+2\hat{q}^2 \mathfrak{q}_1^2\left( e^2+1\right)\right\}+\mathcal{O}\left(p\right)^{-1}.
    \end{align}
\end{subequations}
In the final expression for the PS, these expressions can help us to eliminate the dependence on unnecessary kinetic variables. 
Note there exists other ways to obtain the relation between $(\hat{E},\, \hat{L})$ and $(p,\, e)$. Substituting Eq. \eqref{r12OnP} into Eq. \eqref{eq:r1r2}, we obtain 
\begin{align}
V\left(\frac{p}{1+e}\right)=\hat{L}^2=V\left(\frac{p}{1-e}\right).\end{align} 
Clearly, this is more concise than Eq. \eqref{G0G1}, and we can show that they are essentially equivalent to Eq. \eqref{G0G1}.

Now substituting Eq. \eqref{G0G1} into Eq. \eqref{eq:FuCompare} and further into Eq. \eqref{eq:dudphi}, we find 
\begin{widetext}
\be
\begin{aligned}
    \frac{\dd \phi}{\dd \psi}=&\frac{e}{p}\left[\sum_{n=1}^\infty G_n(\hat{E},\hat{L},p,e,\cos \psi)\sin^{2n-2} \psi\right]^{-\frac{1}{2}}\\
    =&\frac{1}{\sqrt{-f_2}}-\frac{f_3 (e  \cos (\psi)+3)}{2 \sqrt{-f_2} }\frac{1}{p}+\frac{1}{16 \left(-f_2\right){}^{3/2} }\left[3f_3^2 \left(e ^2+18\right)-12 f_2 f_4 \left(e ^2+4\right)\right.\\
    &\left.+ \left(36 f_3^2-32 f_2 f_4\right) e  \cos (\psi)+ \left(3 f_3^2-4 f_2 f_4\right) e ^2 \cos (2 \psi)\right]\frac{1}{p^2}+\mathcal{O}[p^{-3}] \label{dphidY}
\end{aligned}
\ee
\end{widetext}
where in the second step we have substituted the first few $G_n~(n=1,2,\cdots)$ and carried out the large $p$ expansion again. Substituting into Eq. \eqref{eq:alphaVraw} and noting that terms proportional to $\cos(k \psi)~(k=1,2,\cdots)$ will not survive the integration, the result for the PN PS is simply found as 
\be
\begin{aligned}\label{alphaquasi}
    \alpha_{\mathrm{PN}}=&\frac{2 \pi }{\sqrt{-f_2}}+\frac{3 \pi  f_3}{\left(-f_2\right){}^{3/2} p}\\
    &+\frac{3 \pi  \left[f_3^2 \left(e ^2+18\right)-4 f_2 f_4 \left(e ^2+4\right)\right]}{8 \left(-f_2\right){}^{5/2} p^2}+\mathcal{O}\left(p\right)^{-3}.
\end{aligned}
\ee
This result however, is still not a true series of $p$ because of the dependence of $f_n$ on $(\hat{E},\, \hat{L})$ and equivalently on $(p,\, e)$. Substituting Eqs. \eqref{eq:NewTonEL} into $f_n$ in Eqs. 
\eqref{eqs:FuCoefficients} and re-expanding in large $p$, we can eventually obtain a true power series form of $p$ for $\alpha_{\mathrm{PN}}$ 
\be\label{eq:FResult}
\alpha_{\mathrm{PN}}=\sum_{n=1}^{\infty} \lb\sum_{j=0}^{n-1}d_{n,j}e^{2j}\rb \frac{1}{p^n},
\ee
with the coefficient for order $p^{-n}$  a polynomial of order $2(n-1)$ of the eccentricity $e$. The first few coefficients for $d_{n,j}$ are 
\begin{widetext}
\begin{subequations}
    \begin{align}
        d_{1,0}=&\frac{\pi}{ a_1 -2 \hat{q} \mathfrak{q}_1 }\left\{a_1 b_1+a_1 c_1-2 a_1^2+2 a_2+2\hat{q} \left[\left(2a_1-b_1-c_1\right)\mathfrak{q}_1-2\mathfrak{q}_2\right] -2\hat{q}^2\mathfrak{q}_1^2 \right\},\\
        d_{2,0}=&\frac{\pi}{\left( a_1 -2 \hat{q} \mathfrak{q}_1\right)^2} \left\{ \frac{1}{2} a_1^2 b_1 c_1- a_1^3 b_1-\frac{1}{4} a_1^2 b_1^2+ a_1^2 b_2+ a_2 a_1 b_1-4 a_1^3 c_1-2 a_1^2 c_1^2+4 a_1^2 c_2\right. \nn\\
        &\left.+4 a_2 a_1 c_1+5 a_1^4-8 a_2 a_1^2+6 a_3 a_1- a_2^2+\hat{q} \left[ \left(-2 a_1 b_1 c_1+4 a_1^2 b_1+ a_1 b_1^2-4 a_1 b_2\right.\right.\right.\nn\\
        &\left.\left.\left.-2 a_2 b_1+16 a_1^2 c_1+ a_1 c_1^2-16 a_1 c_2-8 a_2 c_1-20 a_1^3+31 a_2 a_1-12 a_3\right)\mathfrak{q}_1+\left(-2 a_1 b_1\right.\right.\right.\nn\\
        &\left.\left.\left.-8 a_1 c_1+9 a_1^2+4 a_2\right)\mathfrak{q}_2-12 a_1\mathfrak{q}_3\right]+\hat{q}^2\left[\left(-5 a_1 b_1-20 a_1 c_1+25 a_1^2-20 a_2 +2 b_1 c_1\right.\right.\right.\nn\\
        &\left.\left.\left.- b_1^2+4 b_2- c_1^2+16 c_2\right)\mathfrak{q}_1^2+\left(-30 a_1  +4 b_1  +16 c_1  \right)\mathfrak{q}_1\mathfrak{q}_2-4 \mathfrak{q}_2^2 \right]+\hat{q}^3\left[\left(-11a_1\right.\right.\right.\nn\\
       &\left.\left.\left. +2b_1+8c_1 \right)\mathfrak{q}_1^3+20\mathfrak{q}_1^2\mathfrak{q}_2\right] +\hat{q}^4\mathfrak{q}_1^4 \right\},\\
       d_{2,1}=&\frac{\pi}{\left( a_1 -2 \hat{q} \mathfrak{q}_1\right)^2} \left\{ -\frac{1}{4} a_1^2 b_1 c_1+\frac{1}{2} a_1^2 b_2-\frac{1}{8}a_1^2 b_1^2+ a_1^3 c_1-\frac{5}{8} a_1^2 c_1^2+\frac{1}{2} a_1^2 c_2- a_2 a_1 c_1 \right.\nn\\
       &\left.+\hat{q} \left[\left( a_1 b_1 c_1+\frac{1}{2} a_1 b_1^2-2 a_1 b_2-4 a_1^2 c_1+\frac{5}{2} a_1 c_1^2-2 a_1 c_2+2 a_2 c_1+ a_2 a_1\right)\mathfrak{q}_1\right.\right.\nn\\
       &\left.\left.+\left(-a_1^2+2a_1c_1\right)\mathfrak{q}_2\right]+ \hat{q}^2\left[\left(5 a_1 c_1- a_1^2-2 a_2- b_1 c_1-\frac{1}{2} b_1^2+2 b_2-20 c_1^2+2 c_2\right)\mathfrak{q}_1^2\right.\right.\nn\\
       &\left.\left.+\left(2a_1-4c_1\right)\mathfrak{q}_1\mathfrak{q}_2\right]+\hat{q}^3\left[\left(3a_1-2c_1\right)\mathfrak{q}_1^3\right]-2\hat{q}^4\mathfrak{q}_1^4\right\}.
    \end{align}
\end{subequations}
\end{widetext}

Before we analyze the implication of result \eqref{eq:FResult} in the next section, we would like to compare the two PS, the $\alpha_{\mathrm{NC}}$ in Eq. \eqref{eq:alphaint} obtained using NC approximation and the above $\alpha_{\mathrm{PN}}$. Clearly, these two PS are only comparable when the orbit is both large and near-circular. 

Under this assumption, we now rewrite the PS $\alpha_{\mathrm{NC}}$ using PN kinetic variables $(p,\, e)$. 
To do this, we need to express the $(\hat{E},~\hat{L},~r_c,~a)$ in terms of $(p,\, e)$ and other spacetime parameters. 
First note that the first two of them, $(\hat{E},\, \hat{L})$, have already been linked to $(p,\, e)$ using Eqs. \eqref{eq:NewTonEL}. For $r_c$, although the defining Eq. \eqref{eq:rc} was not always solvable in closed form, it can always be solved perturbatively for large $p$. Actually, substituting in Eqs. \eqref{eq:NewTonEL} and asymptotic \eqref{eq:abcform2}, the solution to $r_c$ as functions of $(p,\, e)$ can be obtained using the method of undetermined coefficients, as
\be\label{eq:rcOnPe}
\begin{aligned}
    &r_c=\frac{p}{1-e^2}+ \left\{ -a_1^2 c_1+a_1 c_2+a_2 c_1+a_1^3-2 a_2 a_1+a_3\right.\\
    &\left.+\hat{q}\left[\left(2 a_1 c_1  -2 a_1^2   +2 a_2 -2 c_2\right) \mathfrak{q}_1+\left(2 a_1-2 c_1\right) \mathfrak{q}_2-2 \mathfrak{q}_3\right]\right.\\
    &\left.+\hat{q}^2\left[\left(a_1 -c_1\right) \mathfrak{q}_1^2-2 \mathfrak{q}_1 \mathfrak{q}_2\right]\right\}  \frac{e^2}{(a_1-2  \mathfrak{q}_1)p}+\mathcal{O}(p)^{-2}.
\end{aligned}
\ee
Similarly, by substituting Eqs. (\ref{eq:NewTonEL}), (\ref{eq:abcform2}) and \eqref{eq:rcOnPe} into $\hat{L}_c$ in Eq. (\ref{eq:rc}) and further into Eq. \eqref{eq:a}, the small parameter $a$ can also be expressed using $(p,\, e)$ as
\be\label{eq:aOnPe}
\begin{aligned}
    a=&\frac{1}{\sqrt{1-e^2}}-1+\frac{e^2}{\sqrt{1-e^2}\left(a_1-2 \hat{q} \mathfrak{q}_1\right)}\left\{-a_1 c_1+a_1^2-a_2\right.\\
    &\left.+\hat{q}\left[\left(-2 a_1 +2 c_1\right) \mathfrak{q}_1+2 \mathfrak{q}_2\right]+\hat{q}^2\mathfrak{q}_1^2\right\}\frac{1}{p}+\mathcal{O}\left(\frac{1}{p}\right)^2.
\end{aligned}
\ee

Eq. \eqref{eq:rcOnPe} and \eqref{eq:aOnPe} allow us to express all terms that depend on $r_c$ and $a$ explicitly in $\alpha_{\mathrm{NC}}$ in Eq. (\ref{eq:alphaint}) in terms of the asymptotic expansion coefficients $a_n,\,b_n,\,c_n,\,\mathfrak{q}_n$ and kinetic variables $(p,\,e)$. Indeed, the expansion coefficients of the metric and electric potential functions around $r_c$, i.e., $a_{cn},\,b_{cn},\,c_{cn},\,h_{cn}$ can also be expressed using $a_n,\,b_n,\,c_n,\,\mathfrak{q}_n$ and variables $(p,\,e)$. To do this, one only needs to substitute the asymptotic forms \eqref{eq:abcform2} as well as the formula \eqref{eq:rcOnPe} into the expansion \eqref{eq:abcform1} and then further expand these functions for large $p$.
The $a_{cn},\,b_{cn},\,c_{cn},\,h_{cn}$ although can be solved easily, their expressions are too lengthy to present here. 
Finally, substituting all these coefficients, as well as Eq. \eqref{eq:rcOnPe} and \eqref{eq:aOnPe} into Eq. \eqref{eq:alphaint}, we are able to express $\alpha_{\mathrm{NC}}$ also as power series of $1/p$. Not surprisingly, we found that the result agrees with the PN formula \eqref{eq:FResult} perfectly. This agreement shows the correctness of results from both methods. Moreover, for the coefficient of each fixed order of $p^{-n}$, its dependence on the eccentricity $e$ is automatically in a polynomial form, indicating that if the small $a$ expansion in Eq. \eqref{eq:alphaint} is carried out to high enough order, the rewritten PN result to low orders will be valid even for large $e$. 

\section{Applications to charged spacetimes\label{sec:applications}}

In this section, we will proceed to implement the two aforementioned methods to determine PS within several specific spacetimes, and investigate the effect of the electric interaction as well as other spacetime parameters on these shifts. 

\subsection{PS in RN spacetime}\label{subsec:RN}

The RN spacetime is the simplest and cleanest for the analysis of PS of charged particles. However, as we will show next, it still captures the main feature of the electric interaction effect on the PS. The line element and the electric potential of the RN spacetime are given by
\begin{subequations}\label{eqs:RNmetric}
    \begin{align}
&A(r)=\frac{1}{B(r)}=1-\frac{2M}{r}+\frac{Q^2}{r^2},~C(r)=r^2,\\
&\mathcal{A}_0(r)=-\frac{Q}{r}.
    \end{align}
\end{subequations}

We next analyze the  $\alpha_{\mathrm{NC}}$ in Eq. \eqref{eq:alphaint} and $\alpha_{\mathrm{PN}}$ in Eq. \eqref{eq:FResult} in this spacetime. 
First substituting Eqs. \eqref{eqs:RNmetric} into Eq. \eqref{eq:rc}, one can see that $r_c$ for the RN spacetime is determined by a quartic polynomial 
\begin{align}\label{eq:RN_EonRc}
 &r_c^4 \left(\hat{E}^2-1\right)+r_c^3 \left(-3 \hat{E}^2 M-\hat{q}\hat{E}  Q+4  M\right)\nn\\
 &+2 r_c^2 \left( \hat{E}^2 Q^2+2\hat{q} \hat{E} M  Q-2  M^2-  Q^2\right)\nn\\
 &+r_c\left(-3 \hat{q}\hat{E} Q^3+4  M Q^2-\hat{q}^2 M  Q^2\right)+Q^4 \left(\hat{q}^2-1\right)=0 .
\end{align}
Its solution although can still be expressed in a closed form, is too lengthy to show here \cite{Zhou:2022dze}. Once $r_c$ is found, using $\hat{L}_c$ in Eq. \eqref{eq:rc} and Eq. \eqref{eq:a}, one immediate finds the small parameter $a$
\begin{equation}
    a=\frac{r_c}{\hat{L}}\left[\frac{( r_c \hat{E} -\hat{q} Q)^2}{r_c (r_c-2 M)+Q^2}-1\right]^{\frac{1}{2}}-1.
\end{equation}
Using Eqs. \eqref{eqs:tn}, the coefficients $t_n$ for RN spacetime can be found out without much difficulty. After combining with $a$, finally we obtain the $\alpha_{\mathrm{NC}}$ in RN spacetime 
\begin{widetext}
\be
\begin{aligned}\label{eq:RNresult1}
    \alpha_{{\mathrm{RN,NC}}}=&2\pi r_c\left[\frac{( r_c \hat{E} -\hat{q} Q)^2}{r_c (r_c-2 M)+Q^2}-1\right]\left[r_c (r_c-2 M)+Q^2\right] / \left\{   \left(1-\hat{E}^2\right)r_c^6 -6 M  \left(1- \hat{E}^2\right)r_c^5-3  \left(\hat{E}^2-1\right) \left(4 M^2+Q^2\right)r_c^4\right.\\
    &\left.+ \left[-2 Q \left(-8 \hat{E}^2 M Q+ \hat{q}\hat{E} Q^2-4 \hat{q} \hat{E}  M^2+\hat{q}^2 M  Q\right)-8 M^3-6 M Q^2\right]r_c^3
    +3 Q^2 \left[Q \left(-2 \hat{E}^2 Q-4 \hat{q}\hat{E} M +\hat{q}^2 Q\right)\right.\right.\\
    &\left.\left.+ 4 M^2+Q^2\right] r_c^2 +6 Q^4  \left(\hat{q}\hat{E}  Q- M\right)r_c+Q^6 \left(1-\hat{q}^2\right)\right\}^{\frac{1}{2}}-2\pi +\mathcal{O}\left(a\right)^1,
\end{aligned}
\ee
\end{widetext}
Higher orders can also be found but are too lengthy to show here. With the help of a computer algebraic system, we were able to compute this result to the tenth order of $a$. 

To find the PN PS in RN spacetime, all we need to do is to expand the metric functions and electric potential \eqref{eqs:RNmetric} asymptotically, and then substitute the coefficients into Eq. \eqref{eq:FResult}. These steps are very simple and the result to order $p^{-2}$ is 
\begin{widetext}
\begin{align}
    \alpha_{{\mathrm{RN,PN}}}=&\pi  \left( 6 -\hat{Q}^2-6  \hat{q} \hat{Q} +\hat{q}^2 \hat{Q}^2 \right)\frac{M}{(1-\hat{q}\hat{Q} )p}+\frac{\pi }{4 }\left\{\left[6  \left(18+e ^2\right)-2 \hat{Q}^2 \left(24+e ^2\right)-\hat{Q}^4\right]\right.\nn\\
    &-2  \hat{q} \hat{Q} \left[6  \left(18+e ^2\right)- \hat{Q}^2 \left(37+3 e ^2\right)\right]+2  \hat{q}^2 \hat{Q}^2 \left[ \left(66+e ^2\right)-2 \hat{Q}^2 \left(6+e ^2\right)\right]\nn\\
    &\left.-2  \hat{q}^3 \hat{Q}^3 \left(13-3 e ^2\right)+\hat{q}^4 \hat{Q}^4 \left(1-2 e ^2\right)\right\}\left[\frac{M}{ (1-\hat{q}\hat{Q} )p}\right]^2+\mathcal{O}\left(\frac{M}{p}\right)^3, \label{eq:RNresult2}
\end{align}
\end{widetext}
where $\hat{Q}\equiv Q/M$ denotes the charge-to-mass ratio of the central object. As pointed out at the end of the last section, when both $p$ is large and $e$ is small, this PS should coincide with the NC result $\alpha_{\mathrm{RN,NC}}$ in Eq. \eqref{eq:RNresult1}. Indeed, we have converted all $(\hat{E},\, \hat{L})$ in $\alpha_{\mathrm{RN,NC}}$, including those in $r_c$ and $a$ to $(p,\, e)$, and the result agrees perfectly with \eqref{eq:RNresult2}, which is naturally small $e$ expanded. Eq. \eqref{eq:RNresult2} when truncated to first order and set to neutral test particle, agrees with results in Ref. 
\cite{Avalos-Vargas:2011jja,Heydari-Fard:2019pxd,Mak:2018hcn,Gong:2009zzb,Gong:2008zzb,Heydari-Fard:2019kno,Hu:2013eya,Zhang:2022zox,Yang:2019aez}. For nonzero $\hatq$, Eq. \eqref{eq:RNresult2} to the first order agrees with Eq. (28) of Ref.
\cite{Avalos-Vargas:2012jja} and Eq. (42) of Ref. \cite{Yang:2022sco}. Results in Ref. \cite{Bakry:2021smq, Jiang:2014kxz} however can not be recovered from our work. 

\begin{figure}[htp]
    \centering
        \subfigure[]{\includegraphics[width = 0.4\textwidth]{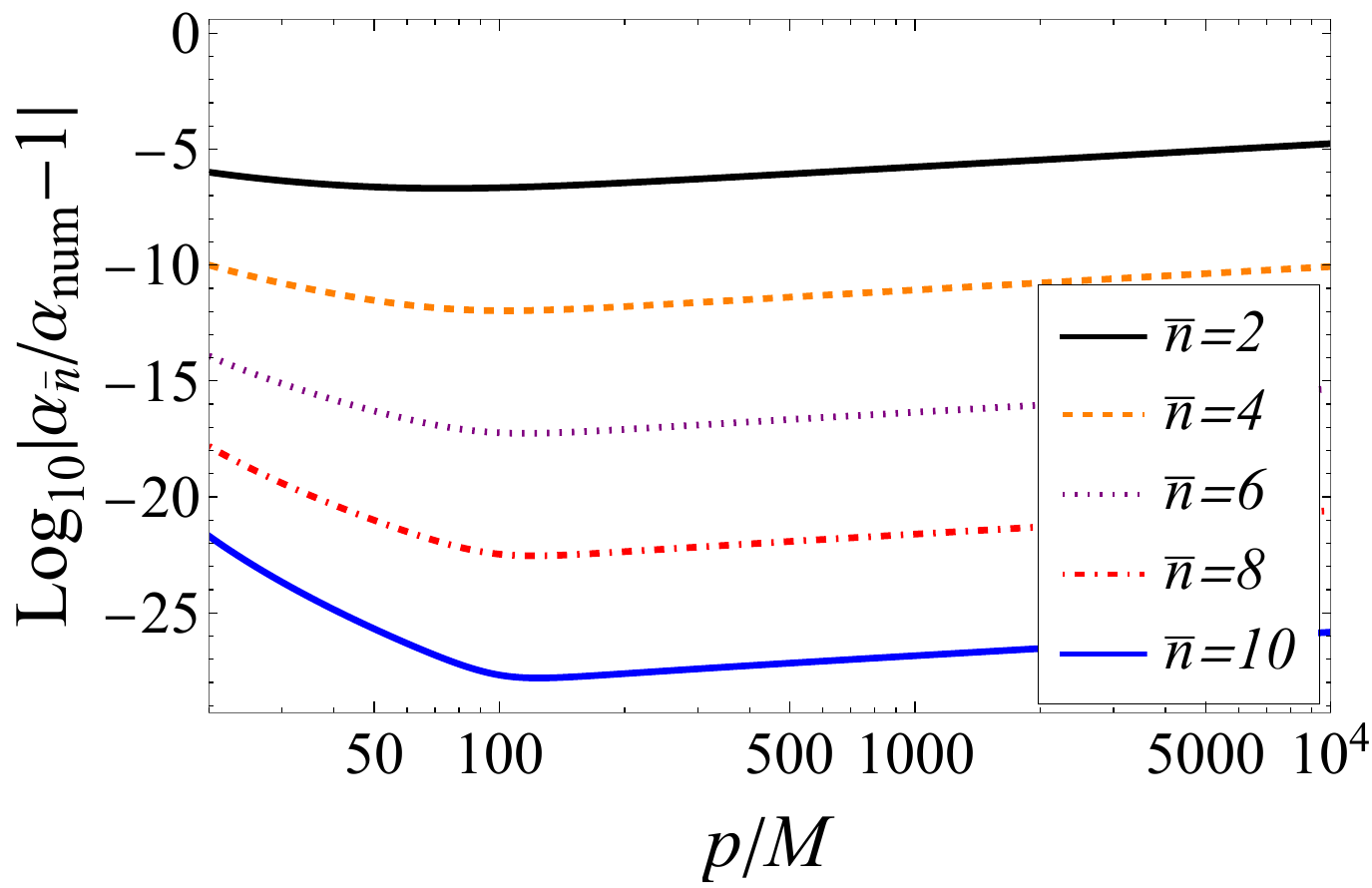}\label{subfig:RNErr_a}}
        \subfigure[]{\includegraphics[width = 0.4\textwidth]{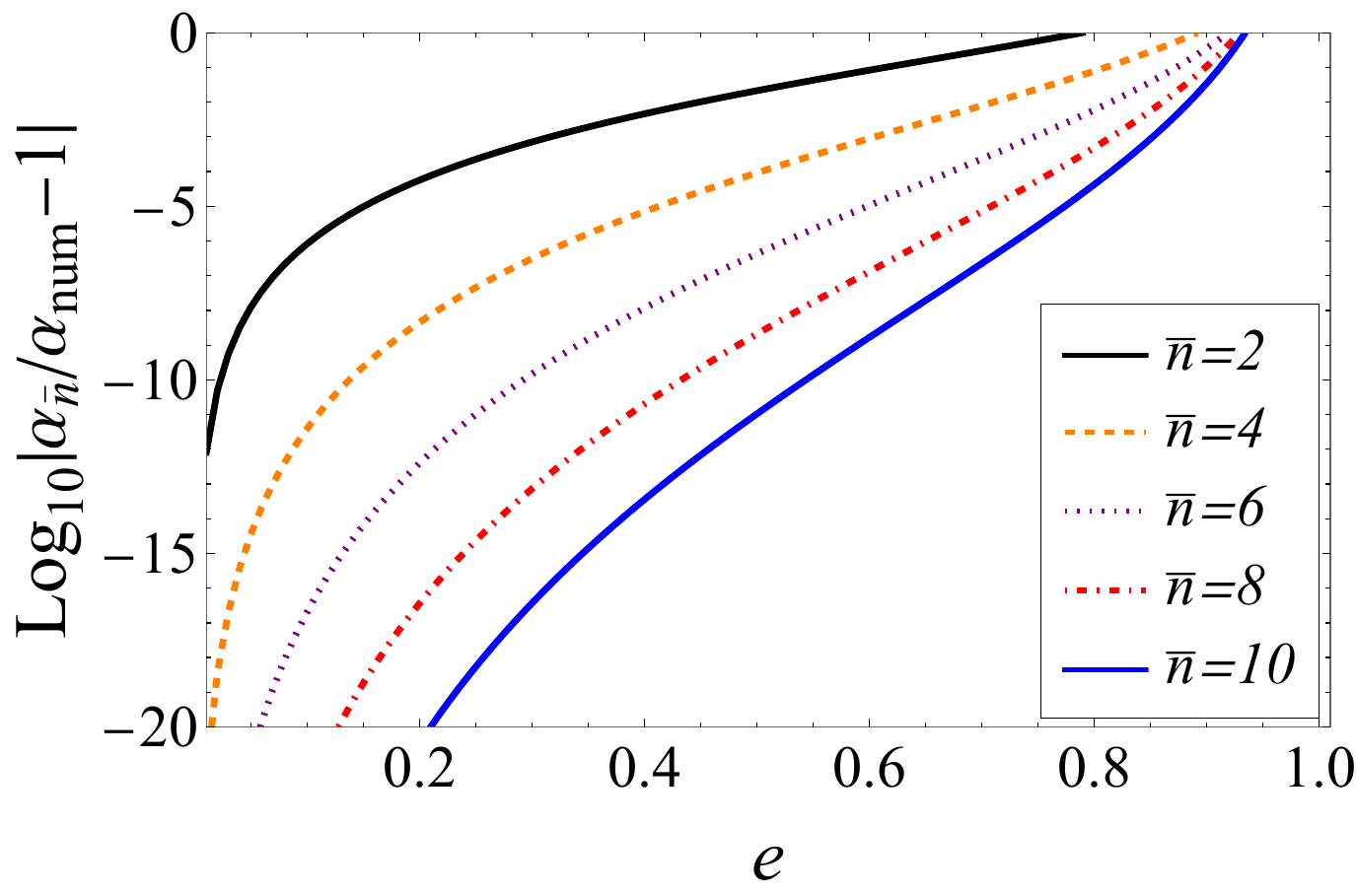}\label{subfig:RNErr_b}}
        \subfigure[]{\includegraphics[width = 0.4\textwidth]{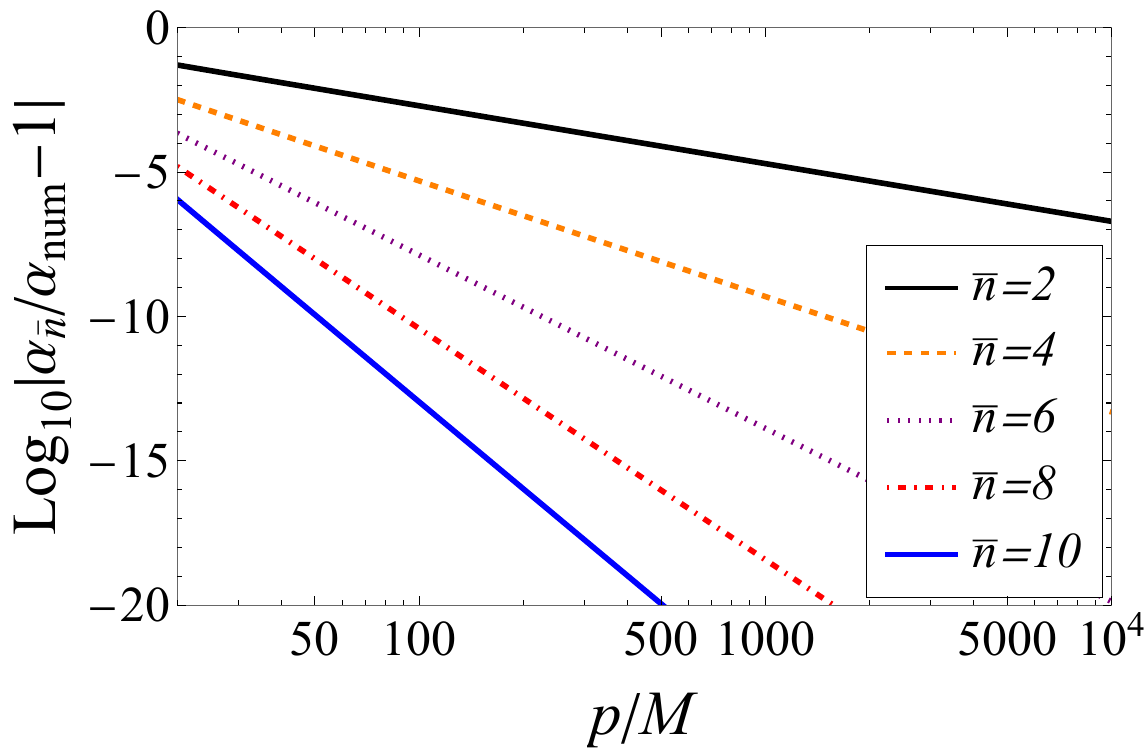}\label{subfig:RNErr_c}}
        \subfigure[]{\includegraphics[width = 0.4\textwidth]{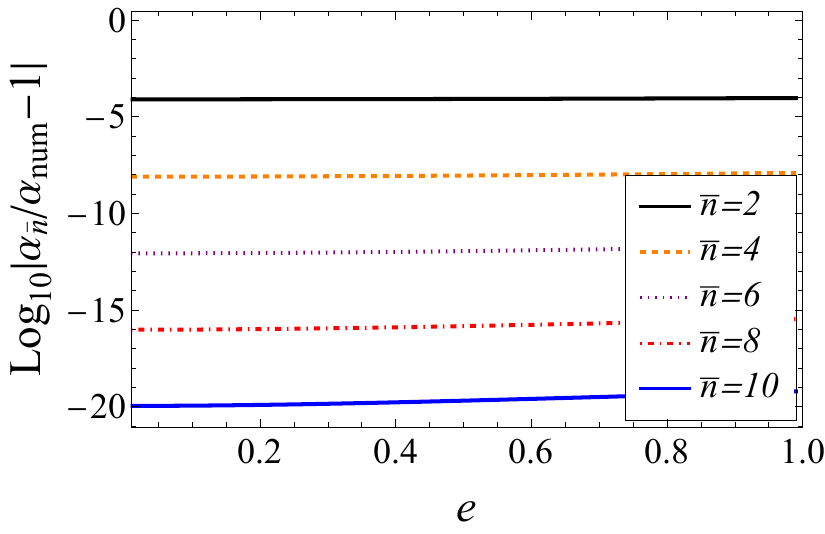}\label{subfig:RNErr_d}}
    \caption{Differences between the numerical result and analytic results Eq. \eqref{eq:RNresult1} and \eqref{eq:RNresult2} truncated to order $\bar{n}$, with parameters fixed at $\hat{Q}=1/2,\hat{q}=-1/5,p=500M,e=1/10$, except the running one.}
    \label{fig:RN_Err_1}
\end{figure}

To check the correctness of the PS \eqref{eq:RNresult1} and \eqref{eq:RNresult2}, in Fig. 
\ref{fig:RN_Err_1} we plot the difference between the series result $\alpha_{\mathrm{RN,NC}}$ (Fig. \ref{subfig:RNErr_a} and \ref{subfig:RNErr_b}) and $\alpha_{\mathrm{RN,PN}}$ (Fig. \ref{subfig:RNErr_c} and \ref{subfig:RNErr_d}) truncated to order $\bar{n}$ and the PS
$\alpha_{\mathrm{num}}$ computed using numerical integration of the definition \eqref{eq:alpha_raw_raw}. The latter is done to very high accuracy and therefore can be regarded as the true value of the PS. In both plots, we fixed $|\hat{Q}\hat{q}|\ll 1$ to make sure that the orbit is a bound one and the PS is well defined. For the NC PS plotted in Fig. \ref{subfig:RNErr_a} and \ref{subfig:RNErr_b}, although $\alpha_{\mathrm{RN,NC}}$ only explicitly depends on $(\hat{E},\, \hat{L})$ we still converted them to $(p,\, e)$ using Eqs. \eqref{eq:NewTonEL} in order to make a comparison with Fig. \ref{subfig:RNErr_c} and \ref{subfig:RNErr_d} for the PN PS $\alpha_{\mathrm{RN,PN}}$. Note that in \ref{subfig:RNErr_a} and \ref{subfig:RNErr_c}, $p$ is varied while in \ref{subfig:RNErr_b} and \ref{subfig:RNErr_d}, $e$ is changed. 

It is seen from all plots of Fig. \ref{fig:RN_Err_1} that as the truncation order increases, both the NC and PN results approach the true value of the PS exponentially. This shows that both methods work well as the series order increases. Moreover, for $\alpha_{\mathrm{RN,NC}}$
Fig. \ref{subfig:RNErr_a} shows that when $e$ is small, the NC approximation almost works equally well for large $p$ and small $p$, and \ref{subfig:RNErr_b} shows that this method is very sensitive to the eccentricity $e$. The smaller the $e$, i.e, the more circular the orbit is, the more accurate the $\alpha_{\mathrm{RN,NC}}$. In contrast, one observes that the features of plots \ref{subfig:RNErr_c} and \ref{subfig:RNErr_d} are the opposite of \ref{subfig:RNErr_a} and \ref{subfig:RNErr_b}: the accuracy of $\alpha_{\mathrm{RN,PN}}$ is not sensitive to $e$ but increases rapidly as $p$ increases. Again, this is in accord with the expectation of a PN approximation of the orbit.
The features mentioned above were also observed in Ref. \cite{He:2023joa} where the effect of the spacetime spin is studied. 

\subsubsection*{Effect of $\hat{q}$ and $\hat{Q}$ on PS}

With the correctness of the PS, especially the PN PS \eqref{eq:RNresult2} verified, we now can use it to analyze the physics contained in this formula.
There are immediately a few points to make. The first is that this PS depends on the parameters $q,~m,~Q,~M$ only through the two charge-to-mass ratios $\hatq$ and $\hatcq$, indicating certain redundancy in these parameters. Therefore in the following when analyzing the effects of these parameters, we only need to concentrate on these two ratios.  The second point to observe is that the charge $q$ only appears in the form of $(qQ)^n$, showing that $q$ influences the PS only through the Coulomb interaction. While $Q$ also manifests in terms that are not a product with $q$, and therefore can affect the PS through the gravitational channel.
The third point is that since this result is a post-Newtonian one, it is more accurate in the asymptotic regions, where gravity follows the universal gravitational law. In such regions, the gravitational attraction and the Coulomb force, which depends on the sign of $qQ$, will strengthen or weaken each other, resulting in a net interaction proportional to $(mM-qQ)$ under the natural unit system we are using. 
Therefore in order for the orbit to be bounded so that a study of the PS is meaningful, this total force has to be an attractive one which means $mM-qQ>0$, i.e, $1-\hat{q}\hat{Q}>0$. This point is reflected in the denominator of each order in Eq. \eqref{eq:RNresult2}. 

In order to fully study the effect of parameters $(m,~q,~Q,~M,~p)$ on the PS using Eq. \eqref{eq:RNresult2}, we have to first determine the boundary of the parameter space in which this series result is convergent. Inspecting Eq. \eqref{eq:RNresult2} more closely, one can observe that this formula contains and only contains power series of the three quantities in the left-hand sides of the inequalities in Eq. \eqref{eq:hatcond}, as well as their product series. In order for the total series to converge, the necessary and sufficient condition is that the sizes of all these three quantities are less than one, i.e.,
\begin{align}
\left[\frac{1}{(1-\hatq\hatcq)}\frac{M}{p}\right]^{1/2}&\leq 1,\nn\\
\left[\frac{\hatcq^2}{(1-\hatq\hatcq)}\frac{M}{p}\right]^{1/2}&\leq 1, \nn\\
\left[\frac{\hatq^2\hatcq^2}{(1-\hatq\hatcq)}\frac{M}{p}\right]^{1/2}&\leq 1.
\label{eq:hatcond}
\end{align}
These conditions can be further simplified to
\begin{align}
1-\hat{q}\hat{Q}\geq \frac{M}{p},~1-\hat{q}\hat{Q}\geq\hat{Q}^2\frac{M}{p},~1-\hat{q}\hat{Q}\geq\hat{q}^2\hat{Q}^2\frac{M}{p}. \label{eq:condinori}
\end{align}

\begin{figure}[htp]
    \centering
\includegraphics[width = 0.45\textwidth]{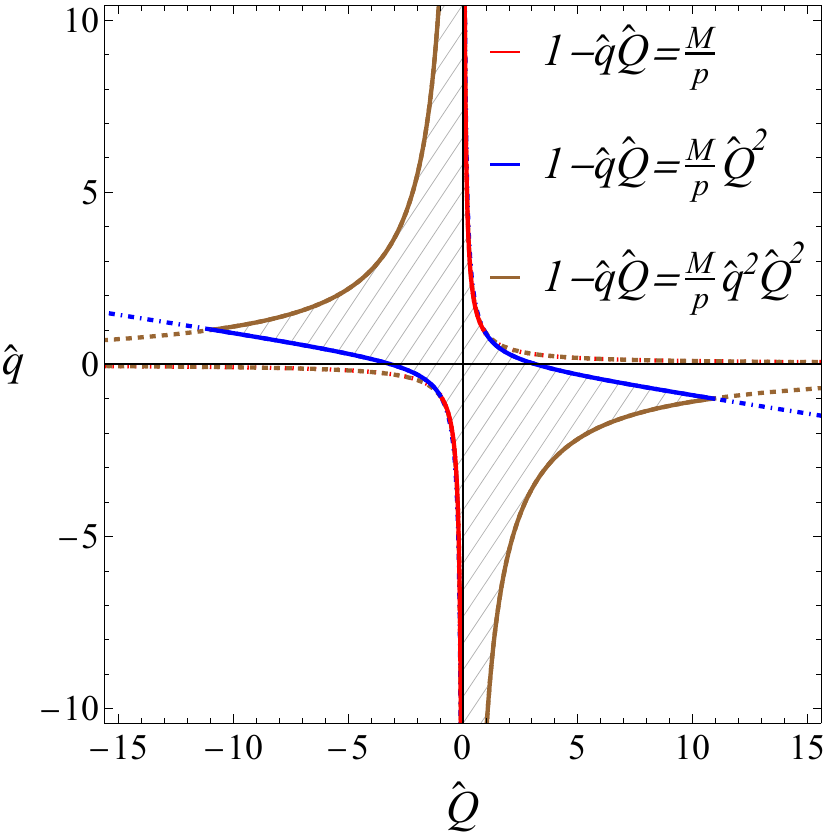}
    \caption{Allowed parameter space for the PN PS $\alpha_{\mathrm{RN,PN}}$ to be valid (gray area). The red, blue and brown boundaries are due to the first, second and third conditions in Eq. \eqref{eq:condinori} respectively. We fixed $p=10M$ in this plot so that the curves are visually separated. A more relativistic value of $p$ will only change this figure quantitatively. }
    \label{fig:parameterspace}
\end{figure}

We have drawn a parameter space spanned by $(\hat{q}, \hat{Q})$ in Fig. \ref{fig:parameterspace} to show the allowed regions bounded by condition \eqref{eq:condinori} and the requirement $1-\hat{q}\hat{Q}>0$. It is seen that this region is mainly concentrated around the $\hat{q}$ and $\hat{Q}$ axes and in the $\hat{Q}$ direction it is completely bounded for a given $M/p$ while in the $\hat{q}$ direction the region is not limited. All three conditions in Eq. \eqref{eq:condinori} are effective in some part of this parameter space. 
In our following studies of the effects of various parameters, we will limit the parameter ranges according to this figure. Note that since the PS depends on $\hatq$ and $\hatcq$ only through $\hatq\hatcq$ and $\hatcq^2$, in principle we can further limit our analysis to the case of $\hatcq\geq0$ for general $\hatq$. The case with $\hatcq<0$ is deducible by switching the sign of $\hatq$ but keeping $\hatcq$'s sign. However, to be as straightforward as possible, we will make the statements in the following applicable to any sign choices of $\hatq$ and $\hatcq$.

\begin{figure}[htp]
    \centering
\includegraphics[width = 0.45\textwidth]{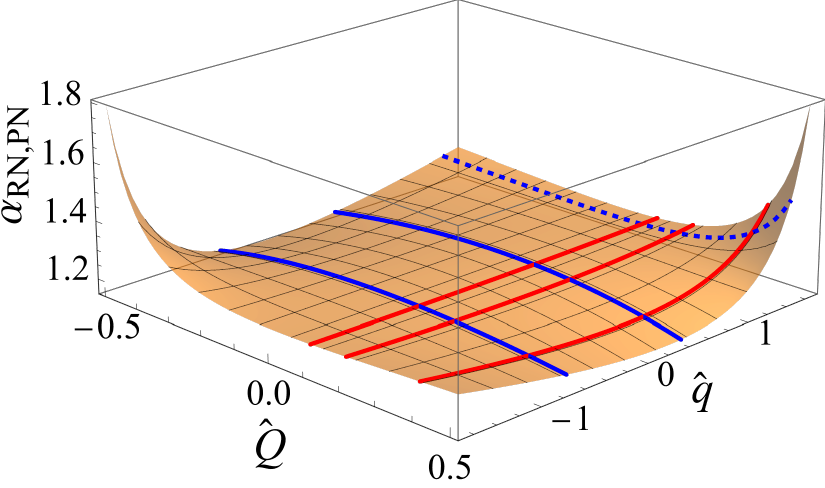}
\caption{Dependence of $\alpha_{\mathrm{RN,PN}}$ for small $\hatq\hatcq$. The red curves correspond to $\hatcq=1/10,\,1/5,\,2/5$ and the solid and dashed blue curves correspond to the $\hat{q}=-7/10,2/5$ and $\hat{q}=3/2$ respectively. We choose $p=20M,~e=4/5,~|\hat{Q}|<2/5,~|\hat{q}|<5/3$ in order for the qualitative features to be visible. }
    \label{fig:rnsmallqq}
\end{figure}

In the following analysis, we examine the influence of variables $\hatq$ and $\hatcq$ on the PS. We will concentrate on the leading order(s) of $M/p$. When $\hatq\hatcq\ll 1$, we can expand the denominator of Eq. \eqref{eq:RNresult2} and the PS to the order $(\hat{q}\hat{Q})^2$ becomes
\be\label{eq:RNQ0q0}
 \begin{aligned}
    \alpha_{\mathrm{RN,PN}}=& \frac{\pi M}{p}\left[6-\hat{Q}^2-\hat{q}\hat{Q}^3 +\left(1-\hat{Q}^2\right) \hat{q}^2\hat{Q}^2 \right]\\
    &+\mathcal{O}\left[\frac{M}{p},(\hat{q}\hat{Q})^3\right].
\end{aligned}
\ee
When $\hatq\hatcq=0$, clearly all electrostatic effect drops out and the Schwarzschild result to the leading order is recovered. For a fixed nonzero $\hatcq$ and $0<|\hatq\hatcq|\ll 1$, it is seen that the first order term $(\hatq\hatcq)^1\hatcq^2$ will dominate the second order $(\hatq\hatcq)^2$ term. Then comparing to neutral particles, a small electrostatic attractive (or repulsive) force due to a small $|\hatq|$ with $\mathrm{sign}(\hatq\hatcq)=-1$ (or $\mathrm{sign}(\hatq\hatcq)=+1$) will increase (or decrease) the PS, similar to the effect of a larger (or smaller) $M$ to the PS of neutral particles in Schwarzschild spacetime. When $|\hatq|$ grows larger than $|\hatcq|/2$ however, we see from Eq. \eqref{eq:RNQ0q0} that the $\hatq^2\hatcq^2$ term becomes larger than $\hatq\hatcq^3$ and therefore the PS will increase as $|\hatq|$ increases. In Fig. \ref{fig:rnsmallqq}, we plot the dependence of the PS on $\hatq$ for three typical $\hatcq$ ($\hatcq=1/10,\,1/5,\,2/5$) using the red curves. For very small $\hatcq$, the dependence is visually suppressed by the largeness of the PS at large $\hatq$ and $\hatcq$. If we fix $\hatq$ nonzero and allow $\hatcq$ to vary but still keep $0<|\hatq\hatcq|\ll 1$, then as $\hatcq$ deviates from zero, 
the PS \eqref{eq:RNQ0q0} to the leading orders of $\hatcq$ will behave as
\be\label{eq:RNQ0q02}
 \begin{aligned}
    \alpha_{\mathrm{RN,PN}}= \frac{\pi M}{p}\left[6-(1-\hatq^2)\hat{Q}^2+\calco(\hatcq)^3\right]+\mathcal{O}\left[\frac{M}{p},(\hat{q}\hat{Q})^3\right].
\end{aligned}
\ee
This means in this region, the increase of $\hatcq$ will decrease (or increase) the PS if $|\hatq|$ is smaller (or larger) than 1. This dependence on $\hatcq$ is shown by the blue curves in Fig. \ref{fig:rnsmallqq}.

\begin{figure}[htp]
    \centering
\includegraphics[width = 0.45\textwidth]{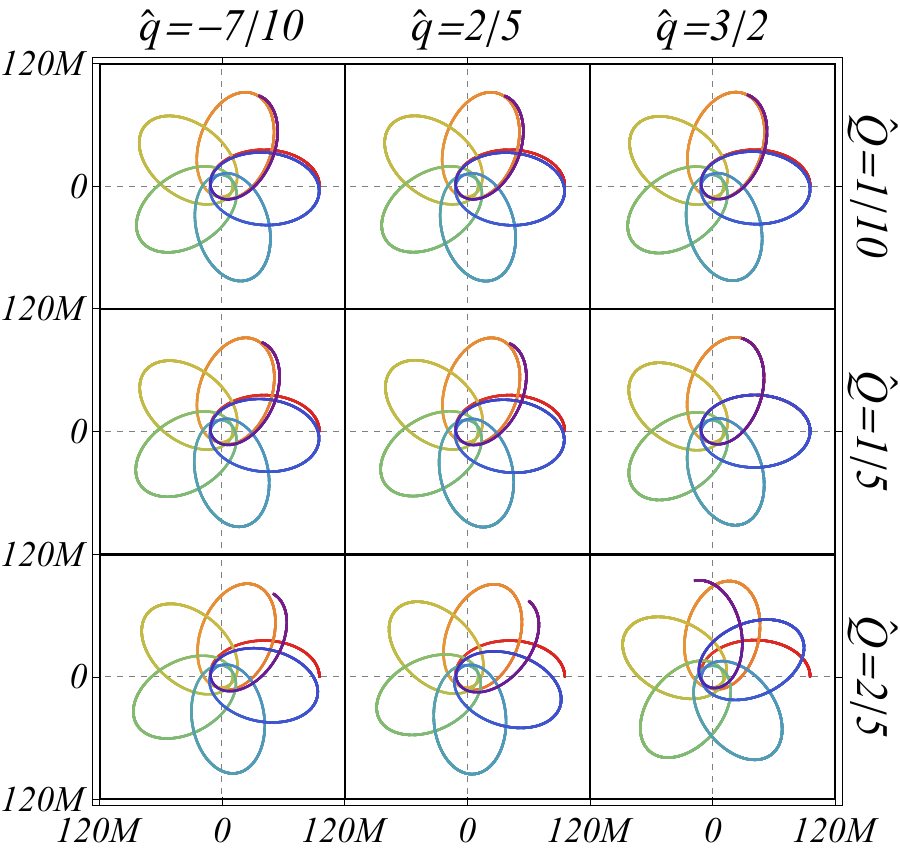}
\caption{The orbits and their PS of charged test particles in RN spacetime. The corresponding $\hatq$ and $\hatcq$ are shown on top and right side of the plots. We choose $p=20M,e=4/5$ in order for the PS to be distinguishable by eyes. }
\label{fig:rnsmallqqorbit}
\end{figure}

In Fig. \ref{fig:rnsmallqqorbit}, we plotted the orbits for several choices of $\hatq$ and $\hatcq$ to illustrate the dependence of the PS on them. The starting points of these orbits are all set at the positive $x$-axis, which are also their apoapses. It is seen that the PS decreases first and then grows as $\hatq$ increases for fixed $\hatcq$. For fixed $|\hatq|<1$ (or $|\hatq|>1$), it decreases (or increases) as $|\hatcq|$ increases. These agree perfectly with the analysis above. 

\subsection{PS in EMD gravity}

The EMD theory stands as a captivating theoretical framework that combines general relativity, electromagnetism, and the dilaton field minimally, providing a useful starting point for unifying all the fundamental interactions. For the static and spherical solution in this theory, its line element and electric field are described by  \cite{Clement:2009ai,Azreg-Ainou:2014gja}
\begin{subequations}
\begin{align}
&\mathrm{d}s^{2}=-f_{+}f_{-}^{\gamma}\mathrm{d}t^{2}+f_{+}^{-1}f_{-}^{-\gamma}\mathrm{d}r^{2}+r^{2}f_{-}^{1-\gamma}\mathrm{d}\Omega^{2},\label{eq:emdmetric}\\
&f_{\pm}=1-\frac{r_{\pm}}{r},\quad\gamma=\frac{1-\eta_{1}\lambda^{2}}{1+\eta_{1}\lambda^{2}},\\
&\mathcal{A}_0(r)=-\frac{Q}{r},
\end{align}
\end{subequations}
where $r_\pm$ are the locations of the outer and inner event horizons respectively and $\eta_1=\pm1$ depending on the dilaton/antidilaton nature of the scalar field. And $\lambda$ is the real (anti)dilaton-Maxwell coupling constant. These parameters can be linked to the mass and charge of the spacetime by
\be
2M=r_+ +\gamma r_-,\quad2Q^2=\eta_2(1+\gamma)r_+r_-
\ee
or equivalently 
\begin{align}\label{eq:EMDrpm}
r_+=&M+ \sqrt{  M^2-\frac{2\eta_2 \gamma  Q^2}{1+\gamma}},\nn\\
r_-=&\frac{1}{\gamma}\lb M- \sqrt{M^2-\frac{2\eta_2 \gamma  Q^2}{1+\gamma}}\rb. 
\end{align}
Here $\eta_2=\pm1$ for Maxwell or anti-Maxwell field respectively. When $\gamma=\eta_2=1$, the metric \eqref{eq:emdmetric} reduces to the normal RN spacetime. 

Substituting the asymptotic expansion coefficients of the metric and the electric potential functions in Eq. \eqref{eq:emdmetric} into Eq. \eqref{eq:FResult}, the PS of charged test objects in the PN approximation in the EMD gravity is found, to the leading order of $1/p$, as 
\begin{widetext}
\begin{align}
    \alpha_{{\mathrm{EMD,PN}}}=&\pi  \left\{ 6 -\eta_2 \hat{Q}^2-  \hat{q} \hat{Q}\left[6+\frac{\gamma-1}{\gamma}\left(1-\sqrt{1-\frac{2\eta_2\gamma \hat{Q}^2}{ \left(\gamma+1\right)}}\right)\right] +\hat{q}^2 \hat{Q}^2 \right\}\frac{M}{(1-\hat{q}\hat{Q} )p}+\mathcal{O}\left(\frac{M}{p}\right)^2. \label{eq:pspnemd}
\end{align}
\end{widetext}
It is seen that the Maxwell-gravity coupling constant $\eta_2$ also determines the sign of its contribution to the PS through gravity (the $\eta_2\hat{Q}^2$ term). This sign choice also affects the PS through the electric interaction term proportional to $\hat{q}\hat{Q}$, but only weakly when $\hat{Q}$ is small. The parameter $\gamma$, which is related to the strength $\lambda$ of the Maxwell-(anti-)dilaton coupling, determines the amount of deviation of the electric interaction from the standard RN case. In the limits $\eta_2=\gamma=1$, this PS agrees with the PS of the RN case in Eq. \eqref{eq:RNresult2}. 

\begin{figure}[htp]
    \centering
\subfigure[]{\includegraphics[width = 0.45\textwidth]{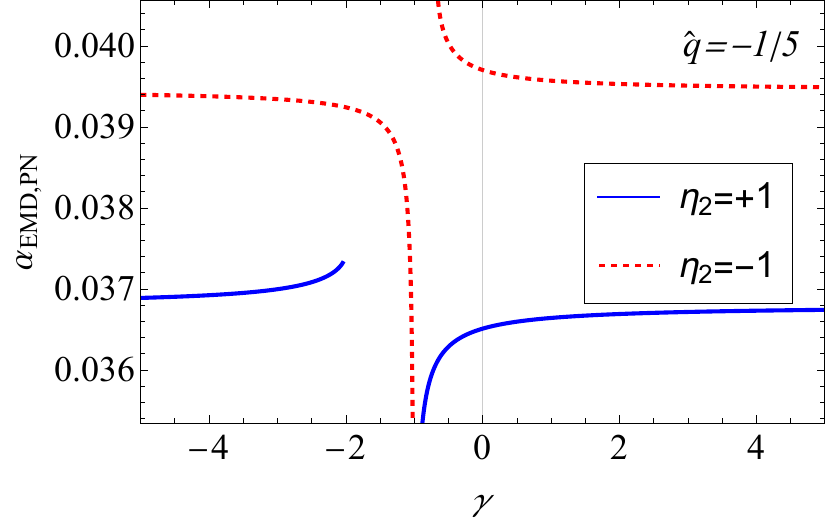}}
\subfigure[]{\includegraphics[width = 0.45\textwidth]{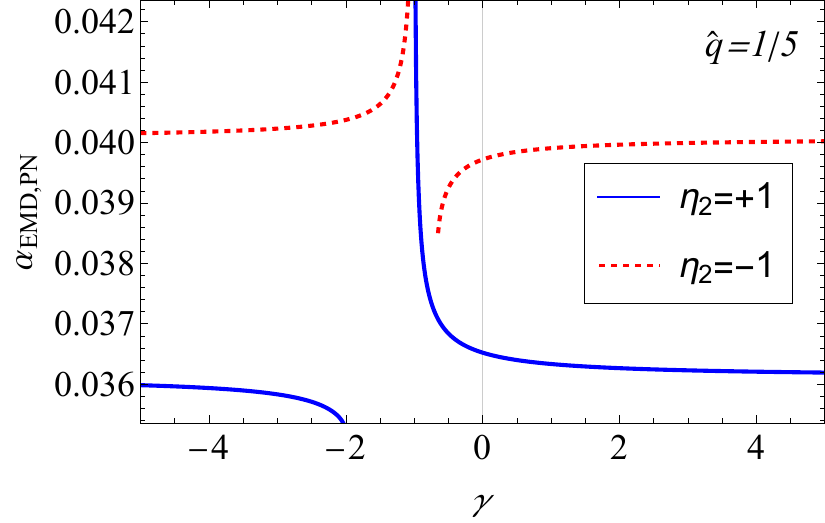}}
\caption{Dependence of $\alpha_{\mathrm{EMD,PN}}$on $\gamma$ with $p=500M,~e=1/10,~\hat{Q}=1/2,~\hat{q}=\pm 1/5$. In order for $r_+,~r_-$ in Eq. (\ref{eq:EMDrpm}) to exist, it is required that $\gamma<-2$ for $\eta_2=+1$ and $\gamma>-2/3$ for $\eta_2=-1$. The blue and red curves correspond to $\eta_2=+1$ and $\eta_2=-1$ respectively. }
    \label{fig:ENDonGamma}
\end{figure}

One can find by inspecting the higher order terms of the PS \eqref{eq:pspnemd} that the coefficient of order $1/p^{-n}$ contains factor of $\eta_2\left(1+\gamma\right)$ up to order $-\frac{n}{2}$, therefore in order for the total PS to converge, the condition $p>1/\sqrt{|1+\gamma|}$ should be fulfilled.  
In Fig. \ref{fig:ENDonGamma}, we show how the Maxwell-(anti)dilaton parameter $\gamma$ affects the PS for different sign choices of $\hat{q}$ and $\eta_2$, with all other continuous parameters fixed. It is seen that for the region of $\gamma$ that the PS are well defined for both $\eta_2$, the effect of $\gamma$ is always opposite to each other for different $\eta_2$ and signs of $\hatq$. 
Near the location of $\gamma=1$, the PS decreases (or increases) as $\gamma$ increases for normal Maxwell field, i.e.,$\eta_2=+1$ (or anti-Maxwell field i.e. $\eta_2=-1$), for positive $\hatq$. This means stronger Maxwell-dilaton coupling will cause a smaller (or larger) PS if the electromagnetic field is Maxwell (or anti-Maxwell) for positive particle charges. In comparison, the effect of $\gamma$ when $\gamma\lesssim 2$ is the opposite to its effect when $\gamma\gtrsim 1$. When $|\gamma|\gg 1$, then its effect on the PS diminishes as $|\gamma|$ increases, as can be seen from Eq. \eqref{eq:pspnemd}.

\subsection{PS in charged wormhole spacetime}

A Lorentzian traversable wormhole with a charge is described by the line element \eqref{eq:gmetric} with metric functions and field strength \cite{Kim:2001ri}
\begin{subequations}
    \begin{align}
A(r)=&\left(1+\frac{Q^2}{r^2}\right),\nn\\
B(r)=&\left(1-\frac{s(r)}{r}+\frac{Q^2}{r^2}\right)^{-1},\label{eq:cwmetric}\\
C(r)=&r^2,\nn\\
\mathcal{F}_{tr}=&-\frac{Q}{r^{2}}\sqrt{A(r)B(r)},\label{eq:cWF}
\end{align}
\end{subequations}
where $s(r)$ is the shape function 
\begin{align}
s(r)=s_{0}^{\frac{2\beta}{2\beta+1}}r^{\frac{1}{2\beta+1}},
\end{align} and $\beta<-1/2$. In order for the PN method to work, the exponent $\frac{2\beta}{2\beta+1}$ has to be an integer and therefore we will fix $\beta=-1$ and consequently $s(r)=s_0^2/r$. 
This implies that $s_0$ plays the role of an additional charge, although only partially, through the metric function $B(r)$. The parameter $s_0$ in this spacetime has to satisfy $s_0^2>Q^2$ to maintain the wormhole. 

The asymptotic form of the metric functions \eqref{eq:cwmetric} can be obtained easily. For the potential $\mathcal{A}_0$, we can get its asymptotic expansion by expanding and then integrating \eqref{eq:cWF}
\begin{align}
    \mathcal{A}_0(r)=-\frac{Q}{r}-\frac{Qs_0^2}{6 r^3}+\frac{Qs_0^2\left(4Q^2-3s_0^2\right)}{40r^5}+\mathcal{O}\left(\frac{Q}{r}\right)^7. \label{eq:cwa0}
\end{align}
Substituting these expansions into Eq. \eqref{eq:FResult}, the PS in this charged wormhole spacetime becomes 
\be
\begin{aligned}
    \alpha_{{\mathrm{CW,PN}}}=&\pi\left(\frac{1}{\hat{q}}-\hat{q}\right)\frac{Q}{p}-\frac{\pi}{4}\left[24+4e^2+\frac{1}{\hat{q}^2}-\hat{q}^2\left(1-2e^2\right)\right.\\
    &\left.-2\frac{s_{0}^{2}}{Q^2}\left(4+e^2\right)\right]\frac{Q^2}{p^2}+\mathcal{O}\left(\frac{Q}{p}\right)^3. \label{eq:cwps}
\end{aligned}
\ee
When $s_0$ is set to zero, this agrees with the PS \eqref{eq:RNresult2} in the RN case with $M=0$. Therefore the PS in this spacetime can be thought of as that of a massless RN spacetime with some modification from parameter $s_0$. And moreover, the fundamental scale in this spacetime can be chosen as the charge $Q$. 

Among the two terms in the leading order, the term proportional to $Q/\hat{q}$ is due to the gravitational effect of the spacetime charge, while the term proportional to $-\hat{q}Q$ is still caused by the electric interaction. The extra parameter $s_0$ only appears starting from the $(Q/p)^2$ order and affects the PS only through gravitation (no multiplication with $\hat{q}$). This also agrees with the fact that $s_0$ does not appear in the leading order of the electric potential in Eq. \eqref{eq:cwa0}. 
Similar to the previous analysis below Eq. \eqref{eq:RNresult2} for the convergence requirement in the RN spacetime, the $\alpha_{{\mathrm{CW,PN}}}$ here also consists of four distinct series and the convergence of the PS requires
\begin{align}\label{eq:cwqQregion}
|Q|<p,~\frac{|Q|}{p}\leq |\hat{q}| \leq \frac{p}{|Q|},~|s_0|<p.
\end{align}
Besides, differentiating Eq. \eqref{eq:motion_r} with respect to $\tau$ again, we find in this spacetime that
\begin{align}
    \ddot{r}=&\frac{\hat{q}Q}{r^4\left(Q^2+r^2\right)}\sqrt{\frac{Q^2+r^2}{Q^2+r^2-s_0^2}}\left[\left(Q^2+r^2\right)^2-Q^2 s_0^2\right]\dot{t}\nn\\
    &+\frac{Q^2 \left(Q^2+r^2-s_0^2\right)\dot{t}^2}{r^5}+\frac{\left(s_0^2-Q^2\right)\dot{r}^2}{r\left(Q^2+r^2-s_0^2\right)}\nn\\
    &+\frac{\left(Q^2+r^2-s_0^2\right)\dot{\phi}^2}{r}.
\end{align}
Since $Q^2+r^2-s_0^2>0$ due to $B(r)>0$ in the observable region, we see from the above equation that if $qQ>0$, $\ddot{r}$ will always be positive and there will exist no bounded orbits or well-defined PS. 
Therefore we will also require that $qQ<0$ in this spacetime in order to study its PS. This is also consistent with the instinct that when there exists no gravitational attraction from the mass ($M=0$), $qQ$ has to be negative in order for the electrostatic interaction to be attractive and to form a bound orbit.  

\begin{figure}[htp]
   \centering
\includegraphics[width = 0.45\textwidth]{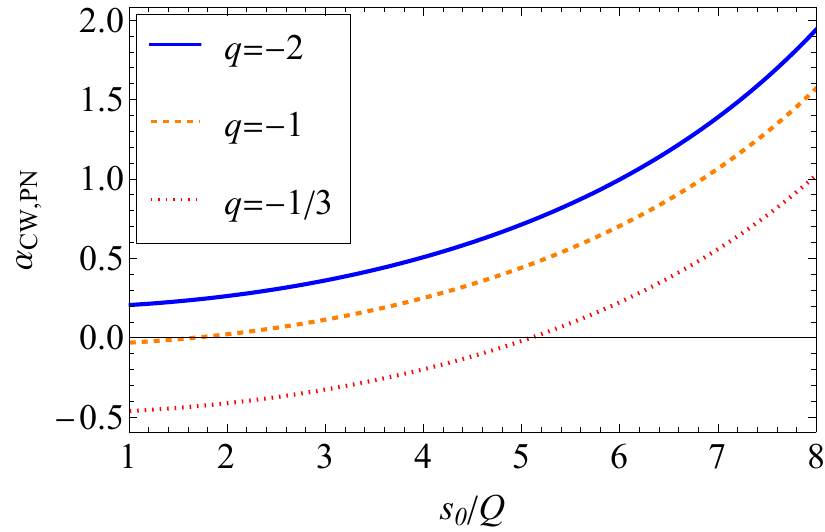}
   \caption{Dependence of $\alpha_{\mathrm{CW,PN}}$ on $s_0$ with fixed $Q=1,~p=20Q,~e=1/2$. } \label{fig:cw_s0}
\end{figure}

In Fig. \ref{fig:cw_s0}, we plot the dependence of the PS on the new parameter $s_0$ (in scale of $Q$) according to Eq. \eqref{eq:cwps} for a few values of $\hatq$. It is seen that for all $\hatq$, the PS increases as $s_0/Q$ increases. This is roughly consistent with the observation in the RN spacetime in Fig. \ref{fig:rnsmallqq} that as the charge $Q^2$ increases the PS also increases. It is also observed that for each $-1\lesssim\hatq<0$, there exists a value of $s_0/Q$ below which the PS becomes negative. This feature however is not present in the regular RN case since there, the PS is mostly positive for reasonable parameter choices.  This critical value of $s_0$ can also be solved from Eq. \eqref{eq:cwps} to the leading two orders as
\begin{align}
s_{0}^2 = \frac{2Q^2}{4+e^2}\left(\hat{q}-\frac{1}{\hat{q}}\right)\frac{p}{Q}+\mathcal{O}\left(Q\right)^2.
\end{align}
From this we see that for a positive $Q$, since $qQ$ has to be negative, roughly the existence of such critical $s_0$ requires that $1\lesssim\hatq<0$, which is consistent with what is observed in Fig. \ref{fig:cw_s0}.

\begin{figure}[htp]
    \centering
\includegraphics[width = 0.45\textwidth]{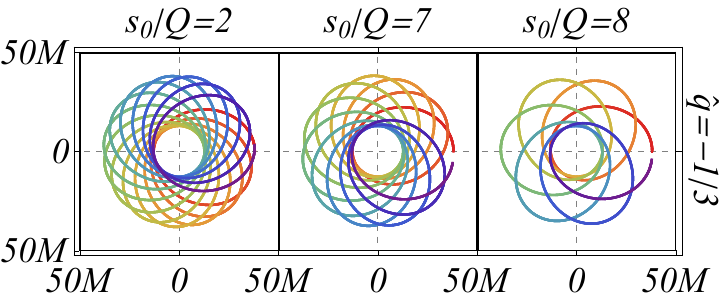}
\caption{Orbits of charged test particles in the charged wormhole. We choose $Q=1,~\hatq=-1/3,~p=20Q,~e=1/2$ and three $s_0/Q$ as indicated in the plot.}
\label{fig:cworbit}
\end{figure}

In Fig. \ref{fig:cworbit}, we choose $\hatq=-1/3$, which corresponds to the red dotted curve in Fig. \ref{fig:cw_s0}, and three typical values of $s_0/Q~( 2, 7, 8)$ to plot the corresponding orbits.
The first of these according to Fig. \ref{fig:cw_s0} corresponds to the orbit with a negative PS while the latter two have positive PS. It is seen from Fig. \ref{fig:cworbit} that these are indeed the case, and we have checked that the PS in these orbits agrees quantitatively with the value in Fig. \ref{fig:cw_s0} and Eq. \eqref{eq:cwps}.

\section{Constraint on charges of known systems \label{sec:appdata}}

In this section, we will apply the PS in the RN spacetime to the observed PS of Mercury around the Sun and S2 around the Sgr A$^*$ to constrain the charges of these objects. 

\subsection{Constraints on solar and Mercury charges}

\begin{figure}[htp]
    \centering
\includegraphics[width = 0.45\textwidth]{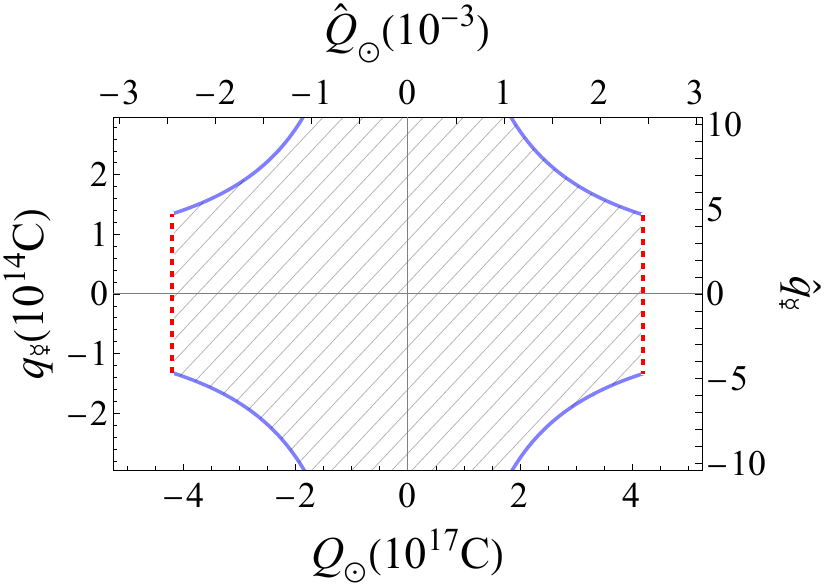}
    \caption{Allowed parameter space of $\hat{q}_{\mercury}$ and $\hatcq_{\odot}$. $M_\odot=1.988\times 10^{30}~\mathrm{kg}$, $M_{\mercury}=3.301\times 10^{23}\mathrm{kg}$ \cite{IAU:2009}, $e=0.2056$, $p=5.545\times 10^{10}~ \mathrm{m}$ \cite{Park:2017} for Mercury are used. }
    \label{fig:mercury}
\end{figure}

The MESSENGER spacecraft measured an uncertainty of $9\times 10^{-4}~''/$cty for the Schwarzschild-like precession of Mercury around the Sun \cite{Park:2017}. If we associate this uncertainty to the electric effect of the solar charge  $\hatcq_\odot$ and the Mercury charge $\hatq_{\mercury}$, then this uncertainty can be used to constrain the charges in the parameter space spanned by them.
The result is shown in Fig. \ref{fig:mercury} where the allowed region of $\hatq_{\mercury}$ and $\hatcq_\odot$ are enclosed by four segments of blue boundaries and two vertical red boundaries. The blue boundaries are due to the leading order of Eq. \eqref{eq:RNresult2} because in this case $p_{\mercury}/M_{\odot}=3.8\times 10^7$ which makes the first order enough for the estimation of the PS. In other words, we treat this uncertainty as
\begin{align}
\Delta\alpha =&\pi  \left( 6 -\hat{Q}^2-6  \hat{q} \hat{Q} +\hat{q}^2 \hat{Q}^2 \right)\frac{M}{(1-\hat{q}\hat{Q} )p}-\frac{6\pi M}{p}\nn\\
=&\left(\hat{q}^2 -1\right)\hat{Q}^2 \frac{\pi M}{(1-\hat{q}\hat{Q} )p},
\label{eq:uncfit}
\end{align}
where in the last term of the first line we deduced the Schwarzschild contribution.
From this we see therefore, these blue boundaries are indeed oddly symmetric, meaning that the change of signs $(\hatq\to-\hatq,\,\hatcq\to -\hatcq)$ will not change the extra PS. 
Moreover, we also see that in general these boundaries are not evenly symmetric, i.e., when only one charge changes sign, the value of $\Delta\alpha$ will change. However in the current case since the extra PS is very small, the total value of the allowed $\hatq\hatcq$ is already very small and this nonsymmetry can not be recognized in the plot by bare eyes. 
For the red boundaries on the solar charge at two ends of $\hatcq_\odot$, they are from the tight constraint in Ref. \cite{Iorio:2012dbo}.

The constrained value of the solar charge is indeed much tighter than the previously reported value using the PS data, if we assume the same charge for Mercury (see Eq. (42) and (43) of Ref. \cite{Avalos-Vargas:2012jja}).

\subsection{Constraints on Sgr A$^{*}$ and S2 charges}

The precession of S2 around Srg A$^{*}$ is measured recently by the Gravity group \cite{GRAVITY:2020gka} to yield a ratio $f$ of the measured value to the standard Schwarschild-precession. A value of $f=1.10\pm 0.19$ was obtained. If we associate the deviation of the central value from 1 to the electric effect of the Sgr A$^*$ and S2 charges, then similar to the case of Mercury precession, we can also use this deviation to find the allowed region in the parameter space of $\hatq_{\mathrm{S2}}$ and $\hatcq_{\mathrm{Sgr\ A}^*}$. 

\begin{figure}[htp]
    \centering
\includegraphics[width = 0.45\textwidth]{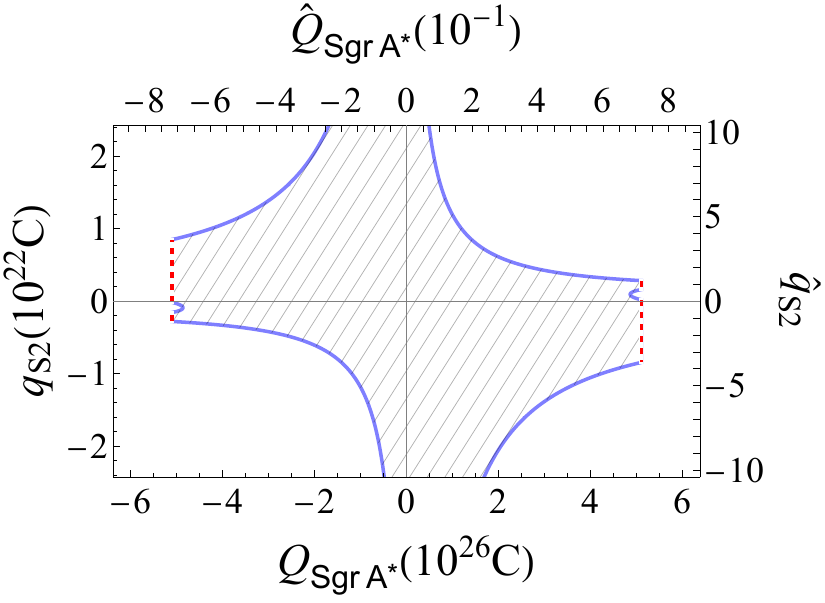}
    \caption{Allowed regions of $\hat{q}_{\mathrm{S2}}$ and $\hatcq_{\mathrm{Sgr\ A}^*}$. $M_{\mathrm{Sgr\ A}^*}=4.15\times 10^6 M_{\odot}$ \cite{2017ApJ...845...22P}, $M_{S2}=13.6M_{\odot}$ \cite{2017Habibi},  $e=0.8843,~p=3.336\times 10^{13}\mathrm{m}$ \cite{2019AA...625L..10G} are used.}
    \label{fig:S2}
\end{figure}

The result is shown in Fig. \ref{fig:S2}, where again the blue boundaries are due to the leading order of Eq. \eqref{eq:RNresult2}. In this case, the $p/M_{\mathrm{Sgr\ A}^*}=5\times 10^3$ \cite{2019AA...625L..10G} is large enough so that the first order of the PS is enough for the estimation of  $\hatq_{\mathrm{S2}}$ and $\hatcq_{\mathrm{Sgr\ A}^*}$. In this allowed parameter space, since the maximal $\hatq\hatcq$ value can be as large as $0.8$, then from Eq. \eqref{eq:uncfit} and comparing to the plot in Fig. \ref{fig:mercury}, its nonsymmetry under sign change $(\hatq\to-\hatq,\,\hatcq\to\hatcq)$ or $(\hatq\to\hatq,\,\hatcq\to-\hatcq)$ is very clear.
The red boundary on the Sgr A$^*$ charge is due to the constraint from the shadow size of the Sgr A$^*$ \cite{Ghosh:2022kit}. To our best knowledge, the allowed region in Fig. \ref{fig:S2} is the first combined constraint on charge $\hatq_\mathrm{S2}$ and $\hatcq_{\mathrm{Sgr\ A}^*}$  using the S2 PS data. 

\section{Conclusions and discussion \label{sec:disc}}

In this work, we employed two methods, the NC approximation and the PN approximation, to systematically study the effect of both gravitational and electrostatic interaction on the PS of charged test particles in general static and spherically symmetric charged spacetimes. The PS using both methods are found to high orders and numerically verified to be very accurate as long as the truncation order is high enough. The NC PS is shown to work better when the eccentricity is smaller while the PN result is more accurate when the semilatus rectum is large, and these results can be shown to be equivalent when both $e$ is small and $p$ is large.

The methods are then applied to the RN, EMD, and charged wormhole spacetimes. It is generally found that the PS only exists for certain regions in the parameter space spanned by the spacetime charge $Q$ and particle charge $q$. Roughly, the condition $qQ<mM$ is always necessary in order for the electrostatic interaction to be weaker than the gravitational attraction if it is repulsive. The spacetime charge $Q$ is found to influence the PS through both the gravitational and electrostatic channels. The combined effect is that the PS will decrease (or increase) if $|\hatq|<1$ (or $|\hatq|>1$) as $|\hatcq|$ deviates from zero. 

The solar-Mercury system and the Sgr A$^*$-S2 system are then modeled using the RN spacetime and their PS data are used to constrain the charge of these objects. For both systems, we found in the $(\hatq,\,\hatcq)$ parameter space the allowed regions of these charges. For the Sgr A$^*$-S2 system, as far as we know, this is the first time that such a constraint is made using the PS data. 

In appendix \ref{sec:psother}, we also present the PS found using the PN method in three other charged spacetimes, the Einstein-Maxwell-scalar gravity, the charged Horndeski and charged black-bounce RN spacetimes. They are not put in the main text because to the first order, their PS (almost) coincide with the RN result and therefore is expected to be non-distinguishable from the RN one using currently available data. 

Although the effect of the electrostatic interaction on the PS in spherically symmetric spacetime is more or less clear. There are still a few possible extensions one can explore. The first is to investigate the effect of a magnetic field on the PS since the magnetic field is (even more) commonly believed to exist in interstellar space. The second is that we can study the effect of other properties of the test particle on the PS, such as when the particle itself is spinning. We are currently pursuing some of these directions. 

\acknowledgements

The authors appreciate the discussion with S. Xu and J. He. This work is partially supported by the Wuhan University Research Development Fund. 

\appendix

\section{PS in other spherical spacetimes\label{sec:psother}}

\subsection{PS with Einstein–Maxwell-scalar field}

For a minimally coupled EMS gravity, the spacetime metric and electric field are described by \cite{Turimov:2021uej}
\be
\begin{aligned}
A(r)=&\frac{1}{B(r)}=\left[\frac{r_+\left(\frac{r-r_-}{r-r_+}\right)^{n/2}-r_-\left(\frac{r-r_+}{r-r_-}\right)^{n/2}}{r_+-r_-}\right]^{-2},\\
C(r)=&A^{-1}(r)\left(r-r_+\right)\left(r-r_-\right),\\
\mathcal{F}_{tr}=&-\frac{Q}{C(r)},
\end{aligned}
\ee
where $n\in(1/2,1]$, and
\begin{align}    r_{\pm}=\frac{M\pm\sqrt{M^{2}-Q^{2}}}{n}
\end{align} are the location of the outer and inner event horizons, and $n$ is related to the scalar charge. 

Substituting the asymptotics of these functions into Eq. \eqref{eq:FResult}, the PS in the EMS field in the PN approximation is found as
\begin{widetext}
\begin{align}
    \alpha_{{\mathrm{EMs,PN}}}=&\pi  \left( 6 -\hat{Q}^2-6  \hat{q} \hat{Q} +\hat{q}^2 \hat{Q}^2 \right)\frac{M}{(1-\hat{q}\hat{Q} )p}+\frac{\pi }{4 n^2}\left[\left(92n^2+24n-8+e^2  \left\{-16n^2+24n-2\right\}\right.\right.\nn\\
    &\left.\left.- \hat{Q}^2 \left\{4\left[13n^2+n-2\right]+2e ^2\left[2n-1\right]\right\}-n^2\hat{Q}^4\right)-2  \hat{q} \hat{Q} \left(92n^2+24n-8 \right.\right.\nn\\
    &\left.\left.-2e^2\left\{8n^2-12n+1\right\}- \hat{Q}^2 \left\{43n^2+2n-8-e^2\left[3n^2+2n-2\right]  \right\}\right)\right.\nn\\
    &\left.+2  \hat{q}^2 \hat{Q}^2 \left(56n^2+14n-4-e^2\left\{12n^2-14n+1-\hat{Q}^2\left[16n^2-4+e^2\left(3n^2-1\right)\right]\right\}  \right)\right.\nn\\
    &\left.-2  \hat{q}^3 \hat{Q}^3 \left(11n^2+2n-e^2\left\{5n^2-2n\right\}\right)+\hat{q}^4 \hat{Q}^4 n^2\left(1-2 e ^2\right)\right]\left[\frac{M}{ (1-\hat{q}\hat{Q} )p}\right]^2+\mathcal{O}\left(\frac{M}{p}\right)^3.
\end{align}
\end{widetext}
To the leading order, this agrees with the RN result \eqref{eq:RNresult2} and therefore it will be difficult to distinguish this spacetime from the RN one using the current observation data.  

\subsection{PS in charged Horndeski spacetime}

The charged Horndeski spacetime is described by the metric and  potential functions \cite{Feng:2015wvb,Cisterna:2014nua}, 
\begin{subequations}
    \begin{align}
        A(r)=&1-\frac{2M}r+\frac{Q^2}{4r^2}-\frac{Q^4}{192r^4},\\
        B(r)=&\frac{1}{A(r)}\left(1-\frac{Q^2}{8r^2}\right)^2,
C(r)=r^2,\label{eq:Horndeski_Ar}\\
\mathcal{A}_0(r)=&-\frac {Q}{r}+\frac{Q^3}{24r^3}.
    \end{align}
\end{subequations}

The expansion of these functions are easy and then substituting into Eq. \eqref{eq:FResult}, the PN PS in this spacetime is found to be
\be
\begin{aligned}
    &\alpha_{{\mathrm{CH,PN}}}=\frac{\pi}{4}  \left( 24 -\hat{Q}^2-24  \hat{q} \hat{Q} +4\hat{q}^2 \hat{Q}^2 \right)\frac{M}{\left(1-\hat{q}\hat{Q} \right)p} \\
    &+\frac{\pi }{64 }\left\{\left[96  \left(18+e ^2\right)-16 \hat{Q}^2 \left(13+e ^2\right)-\hat{Q}^4\right]\right.\\
    &\left.-8  \hat{q} \hat{Q} \left[24  \left(18+e ^2\right)- \hat{Q}^2 \left(43+5 e ^2\right)\right] \right.\\
    &\left.+8  \hat{q}^2 \hat{Q}^2 \left[4\left(66+e ^2\right)- \hat{Q}^2 \left(16+3e ^2\right)\right]-32  \hat{q}^3 \hat{Q}^3 \left(13-3 e ^2\right)\right.\\
    &\left.+16\hat{q}^4 \hat{Q}^4 \left(1-2 e ^2\right)    \right\}\left[\frac{M}{ (1-\hat{q}\hat{Q} )p}\right]^2+\mathcal{O}\left(\frac{M}{p}\right)^3 .
\end{aligned}
\ee
To the leading order, this is different from the RN result in the gravitational contribution of $\hat{Q}$, by a factor of a quarter.

\subsection{PS in charged black-bounce RN spacetimes}

The charge black-bounce RN spacetime is described by the RN metric functions with $r$ replaced by $\sqrt{r^2+l^2}$ where $l$ is some length scale (usually associated with the Planck length), i.e., the new metric functions and electric potential become \cite{Franzin:2021vnj}
\begin{subequations}
    \begin{align}
        A(r)=&\frac{1}{B(r)}=1-\frac{2M}{\sqrt{r^2+l^2}}+\frac{Q^2}{r^2+l^2},\\
        C(r)=&r^2+l^2,\\
        \mathcal{A}_0(r)=&-\frac{Q}{\sqrt{r^2+l^2}}.
    \end{align}
\end{subequations}

The asymptotic expansions of these functions are also simple, and after substituting the coefficients into Eq. \eqref{eq:FResult} we can directly obtain the PS in this spacetime. To the second order, this is
\begin{align}
    \alpha_{{\mathrm{BB,PN}}}=&\alpha_{{\mathrm{RN,PN}}}+\frac{\pi \left(2+e^2\right)l^2}{2p^2}+\mathcal{O}\left(\frac{M}{p}\right)^3,\label{eq:bbresult}
\end{align}
where the $\alpha_{{\mathrm{RN,PN}}}$ is the PS of the RN spacetime in Eq. \eqref{eq:RNresult2}. It is seen that the extra scale parameter $l$ only appears from the second order of $1/p$ and therefore the PS in this spacetime is also difficult to distinguish from that in the regular RN spacetime. Moreover, to the $\mathcal{O}(M/p)^2$ order, the scale parameter $l$  does not participate in any electric interaction effect to the PS. We also checked that this will change starting from the $\mathcal{O}(M/p)^3$ order, i.e., there exist terms proportional to $l^2\hatq\hatcq$ and $l^2\hatq^2\hatcq^2$ in the PS at this order. The result \eqref{eq:bbresult}, after setting $\hatq=0$, reduces to the PS of neutral particles in
black-bounce RN spacetime \cite{Zhang:2022zox}.

\end{document}